\documentclass[10pt,a4paper]{article} 
\usepackage{jcappub}
\usepackage{bm}
\usepackage{comment}
\pdfoutput=0

\newcommand{\up}[1]{{\rm #1}}
\newcommand{\bdi}[1]{\hbox{\boldmath{$#1$}}}
\newcommand{\nnn}{\nonumber \\}
\newcommand{\beeq}{\begin{equation}}
\newcommand{\eneq}{\end{equation}}
\newcommand{\bear}{\begin{eqnarray}}
\newcommand{\enar}{\end{eqnarray}}
\newcommand{\gbar}{\bar g}      
\newcommand{\rbar}{\bar r}      
\newcommand{\UU}{u}             
\newcommand{\VV}{V}    
\newcommand{\VP}{V_\parallel} 
\newcommand{\nP}{n_\perp}     
\newcommand{\zz}{z}           
\newcommand{\ttt}{\theta}     
\newcommand{\pp}{\phi}        
\newcommand{\nhat}{\bdi{n}}
\newcommand{\CK}{\hat k}       
\newcommand{\CT}{\hat\theta}   
\newcommand{\CU}{\hat u}       
\newcommand{\oo}{\Lambda}      
\newcommand{\cc}{\lambda}      
\newcommand{\dea}{{\delta n}}  
\newcommand{\dnu}{\delta\nu}   
\newcommand{\NN}{N}            
\newcommand{\CNN}{\hat N}      
\newcommand{\NC}{\mathbb{C}}   
\newcommand{\dhnu}{\widehat{\Delta\nu}}
\newcommand{\drr}{\delta r}    
\newcommand{\dtt}{\delta\ttt}  
\newcommand{\dpp}{\delta\pp}   
\newcommand{\dT}{\delta\eta}   
\newcommand{\dz}{\delta z}     
\newcommand{\HH}{\mathcal{H}}  
\newcommand{\PP}{\mathcal{P}}
\newcommand{\Dcc}{\Delta\cc}   
\newcommand{\DT}{\Delta\eta}   
\newcommand{\DX}{\Delta x}     
\newcommand{\dL}{\mathcal{D}_L}      
\newcommand{\ddL}{\delta\mathcal{D}_L}
\newcommand{\dA}{\mathcal{D}_A}      
\newcommand{\DD}{\mathbb{D}}         
\newcommand{\dg}{\delta g}           
\newcommand{\GG}{\delta\Gamma}  
\newcommand{\dAp}{d{\cal A}}
\newcommand{\II}{\mathfrak{I}}
\newcommand{\JB}{\mathfrak{B}}
\newcommand{\JR}{\mathfrak{R}}
\newcommand{\JD}{\mathfrak{D}}
\newcommand{\hxi}{\hat\xi}
\newcommand{\htet}{\hat e}
\newcommand{\XX}{\hat\JR}
\newcommand{\OO}{\mathcal{O}}
\newcommand{\ww}{w}
\newcommand{\TT}{\tilde\theta}
\newcommand{\RA}{\rightarrow}

\begin{document}

\begin{titlepage}

\setcounter{page}{1} \baselineskip=15.5pt \thispagestyle{empty}

\bigskip

\vspace{1cm}
\begin{center}
{\fontsize{20}{28}\selectfont \bfseries 
Unified Treatment of the Luminosity Distance in Cosmology}
\end{center}

\vspace{0.2cm}

\begin{center}
{\fontsize{13}{30}\selectfont Jaiyul Yoo$^{a,b}$} ~~and~~
{\fontsize{13}{30}\selectfont Fulvio Scaccabarozzi$^{a}$}
\end{center}

\begin{center}
\vskip 8pt
\textsl{$^a$ Center for Theoretical Astrophysics and Cosmology,
Institute for Computational Science, University of Z\"urich}

\vskip 7pt

\textsl{$^b$Physics Institute, University of Z\"urich,
Winterthurerstrasse 190, CH-8057, Z\"urich, Switzerland}
\end{center}

\note{jyoo@physik.uzh.ch}

\vspace{1.2cm}
\hrule \vspace{0.3cm}
\noindent {\sffamily \bfseries Abstract} \\[0.1cm]
Comparing the luminosity distance measurements to its theoretical predictions 
is one of the cornerstones in establishing the modern cosmology. However, 
as shown in Biern \& Yoo, its theoretical predictions in literature
are often plagued with infrared divergences and gauge-dependences. This trend 
calls into question the sanity of the methods used to derive
the luminosity distance. Here we critically
investigate four different methods ---
the geometric approach, the Sachs approach, the Jacobi mapping approach,
and the geodesic light cone (GLC) approach to modeling
the luminosity distance, and we
present a unified treatment of such methods, facilitating the comparison among
the methods and checking their sanity.
All of these four methods, if exercised properly,
can  be used to reproduce the correct description of the luminosity distance.
\vskip 10pt
\hrule

\vspace{0.6cm}
\end{titlepage}

\noindent\hrulefill

\tableofcontents

\noindent\hrulefill

\section{Introduction}
\label{sec:intro}
Measurements of the luminosity distance from distant supernovas
provided one of the most important
evidence for the mysterious substance in the Universe, or dark energy
\cite{PEADET99,RIFIET98}. However, 
it has been well known that the standard procedure in supernova cosmology
ignores the effect of inhomogeneities in the Universe, 
in comparing the observations to the theoretical predictions
of the luminosity distance. Accordingly, there have been 
numerous theoretical work to account for the inhomogeneities in the Universe
in predicting the luminosity distance.

Within the cosmological framework,
the luminosity distance in an inhomogeneous universe was first computed 
in its complete form \cite{SASAK87} by using the optical scalar equation,
and an explicit check of gauge invariance was also made in \cite{SASAK87}. 
Since this pioneering
work, other methods such as the Jacobi mapping \cite{BODUGA06}, the geometric 
approach \cite{JESCHI12,YOO14a,YOZA14}, the geodesic light cone (GLC) approach 
\cite{GAMAET11} were employed in computing the luminosity distance,
and the calculations were already extended to the second order in 
perturbations, using the optical scalar equation 
\cite{BAMARI05,UMCLMA14a,UMCLMA14b}, the geometric approach \cite{YOZA14}
and the GLC approach 
\cite{BEMAET12,BEGAET13,BEGAET13a,BEDUET14,CLUMET14}. Such second-order
calculations are needed to compute the leading-order corrections to the
mean of the luminosity distance,
in addition to computing its variance. However, these second-order
calculations are complicated by nature, and it is often difficult to compare
the results of calculations by two groups even with the identical approach
adopted. Furthermore, Biern \& Yoo \cite{BIYO16}
showed that most of the luminosity distance calculations in literature, 
even at the linear order 
in perturbations, are often plagued with infrared divergences, 
as the gauge-invariance of its expression is broken in those calculations.

To the linear order in perturbations, at least, this situation makes little
sense, as the full gauge-invariance of the luminosity distance calculation
was explicitly verified \cite{SASAK87,YOFIZA09,YOO10,JESCHI12,YOO14a} 
under a general coordinate transformation. However, when numerical
calculations are performed in practice, a certain 
gauge condition is adopted to facilitate the computation, and several terms
through the process
are ignored in the calculations, breaking the gauge-invariance of the 
expression. Consequently, some sick features in theory arise, such as the 
infrared
divergences of the variance in the luminosity distance \cite{BAMARI05}.
Often this pathology is avoided in most numerical calculations by introducing
a cut-off scale with little justification. 
However, as demonstrated in \cite{BIYO16}, the correct
gauge-invariant calculation of the luminosity distance is devoid of such
pathology, and other numerical errors in the luminosity distance can be
avoided by ensuring the gauge-invariance of its expression.

In comparison to the Planck result \cite{PLANCK13},
recent measurements \cite{RIMAET16} of the luminosity distance and hence the 
Hubble parameter show
signs of discrepancies, and this conflict has renewed interest in the 
luminosity distance calculations, accounting for the inhomogeneities in
the Universe (e.g., \cite{BEGAET13,FLDUUZ13a,BEDUET14}). However, the status
of the second-order calculations of the luminosity distance is even worse
than the linear-order case, as the calculations are much more involved and
an explicit check of its gauge invariance is implausible. While there exist
several groups in literature that have performed the second-order calculations,
there is practically little hope that these calculations can be compared
in a meaningful way to reach a consensus, as their notations and approaches
are vastly different.

Here we provide a unified treatment of the luminosity distance in cosmology,
comparing four different approaches in literature and critically 
assessing the sanity of the methods. This work will serve as a first step 
to go beyond the linear order in computing the luminosity distance and
to build a coherent theoretical framework for all practitioners.
The organization of the paper is as follows. We introduce our notation
convention and present the distortion of a photon path from a straight line
in Section~\ref{sec:formalism}. The distortion of the photon path is further
decomposed into the radial and the angular deviations, and the observed
redshift is related to the time coordinate of the source and its residual
deviation. In Section~\ref{sec:geom}, we use these geometric deviations 
to express the fluctuation in the luminosity distance.
In Section~\ref{sachs}, the optical scalar equation is derived to model the
luminosity distance, and its relation in the conformally transformed metric
is carefully derived. With such relations, the work by Sasaki \cite{SASAK87}
and Umeh et~al. \cite{UMCLMA14a,UMCLMA14b} is derived
with proper corrections and is then related to our geometric approach.
In Section~\ref{Jacobi}, the geodesic deviation equation is presented, and
its counterpart in the conformally transformed metric is derived.
We use the Jacobi mapping approach to derive the fluctuation in the luminosity
distance and compare the result to the work in Bonvin et~al.
\cite{BODUGA06}. In Section~\ref{GLC}, we present the basics of the
geodesic light cone approach to the luminosity distance and provide
corrections to its boundary condition. The luminosity distance with such
corrections is consistent with that from other approaches. Finally, 
we conclude and discuss the implications in Section~\ref{sec:discussion}.

In this paper, we use the indices~$a,b,c,d,\cdots$ to represent the
four-dimensional spacetime indicies, while we use
indicies~$i,j,k,\cdots$ to represent the three-dimensional space indicies.
In certain cases, we also use indices~$I,J,K,\cdots$ in capital letters
to represent the angular indicies in a spherical coordinate.

\section{Light Propagation in Curved Spacetime}
\label{sec:formalism}
In this section, we present the metric convention and derive the propagation
of light in a FRW universe. The real position of the source is decomposed
into the apparent position inferred by the observed redshift and angular 
position
and the residual perturbations in the position, providing basic ingredients
for the geometric approach in Sec.~\ref{sec:geom}.

\subsection{Metric Convention and Photon Wavevector}
\label{ssec:metric}
Being an observable quantity, the luminosity distance is a gauge-invariant
quantity, and it can be computed in any choice of gauge conditions. However,
its computations in literature are predominantly performed in the conformal
Newtonian gauge. To facilitate the comparison among various methods,
we hereafter adopt the conformal Newtonian gauge for our metric representation:
\beeq
ds^2=g_{ab}dx^adx^b=-a^2(1+2\psi)d\eta^2+a^2(1+2\phi)\gbar_{ij}dx^idx^j~,
\eneq
where the conformal time coordinate is~$\eta$, the expansion scale factor 
is~$a(\eta)$, and $\gbar_{ij}$ is the 3-spatial metric tensor in 
the background. Throughout the paper, we will consider two scalar 
degrees~$\psi$ and~$\phi$ of freedom at the linear order in perturbations,
ignoring the vector and tensor perturbations in the metric.
As we are concerned with light propagation ($ds^2=0$),  
it is convenient to consider a conformal transformation:
\beeq
\label{metric}
ds^2=g_{ab}dx^adx^b=a^2\hat g_{ab}dx^adx^b~,
\eneq
where we used a hat to denote quantities in the conformally transformed 
metric~$\hat g_{ab}$. The coordinates $x^a$ 
of photon paths are identical in a given geometry, regardless
of whether the physical metric~$g_{ab}$ or the conformally transformed
metric~$\hat g_{ab}$ is used, greatly simplifying the calculations when
the latter is used.
The four velocity~$u^a$ of timelike flows ($-1=u^au_a$) can be parametrized
to the linear order in perturbations as
\beeq
\UU^a={dx^a\over dt}=\frac1a\left(1-\psi~,~\VV^i\right)~,\qquad
\UU_a=-a\left(1+\psi~,-\VV_i\right)~,
\eneq
where the proper time of the flows is~$dt$ and 3-spatial velocity~$\VV^i$
is based on 3-spatial metric~$\gbar_{ij}$. With respect to the conformally
transformed metric, the four velocity of the timelike flows is then
\beeq
\CU^a=a\UU^a=\left(1-\psi~,~\VV^i\right)~,\qquad 
\CU_a=\hat g_{ab}\CU^b={\UU_a\over a}=-\left(1+\psi~,-\VV_i\right)~.
\eneq

The light propagation is described by two key observable quantities
measured by the observer in the rest frame: 
its phase~$\vartheta$ and propagation direction~$\nhat$. Given these
two observables, the light cone in spacetime is defined as the two-dimensional
surface with a constant phase~$\vartheta$, and the propagation 
direction~$\bdi{k}$ is orthogonal to the constant hypersurface of phase:
\beeq
\label{phase}
\vartheta=\bdi{k}\cdot\bdi{x}-\omega t~,\qquad\qquad
k^a_L=\eta^{ab}\vartheta_{,b}=\left(\omega~,~\bdi{k}\right)~,
\eneq
where the angular frequency is $\omega=2\pi\nu$ in relation to its
wavelength $\lambda=1/\nu$ and we used the subscript~$L$ to emphasize that
the four vector is written in the local rest frame with Minkowsky 
metric~$\eta_{ab}$. For later convenience, we will use the observed
direction of the light propagation $\nhat\equiv -\bdi{k}/|\bdi{k}|$
(opposite to the propagation direction), where the null condition imposes 
$\omega=|\bdi{k}|$. The photon wavevector expressed in terms of local
observable quantities can be transformed to that in a FRW coordinate:
\beeq
\label{wavephy}
k^a={dx^a\over dx_L^b}k_L^b=[e_b]^ak_L^b=
{\omega\over a}\left(1-\psi-\VP~,-n^i+\VV^i+\phi~n^i\right)~,
\eneq
where we defined the line-of-sight velocity~$\VP=\VV^in_i$ and the orthonormal
local tetrads for the metric transformation are
\beeq
[e_t]^a=u^a=\frac1a\left(1-\psi~,~\VV^i\right)~,\qquad\qquad
[e_j]^a={1\over a}\left[\VV_j~,~\delta^i_j(1-\phi)\right]~,
\eneq
defining the proper time-direction and three spacelike four vectors
of the observer in the FRW frame. The local tetrads were constructed
by using the orthonormality condition
($\eta_{ab}=g_{ab}[e_c]^a[e_d]^b$, $c,d=t,x,y,z$). The photon wavevector~$k^a$
in the FRW frame is different from that~$k^a_L$ 
measured in the local rest frame,
since the observer is moving and the gravitational redshift affects the
photon energy.

For the computational convenience, the photon wavevector is again conformally
transformed as
\beeq
\CK^a={dx^a\over d\cc}=\NC a^2k^a~,
\eneq
where the photon path is parametrized by the affine parameter~$\cc$ and
the overall constant factor~$\NC$ reflects the multiplicative freedom
in the parametrization.\footnote{The photon wavevector in 
Eq.~\eqref{wavephy} has no such degree of freedom, since it is completely
specified in terms of physical quantities. Only when conformally transformed,
the wavevector~$\CK^a$ has an additional freedom.} With the expression in 
Eq.~\eqref{wavephy}, the conformally transformed wavevector is
\beeq
\label{normal}
\CK^a=\left(\NC\omega a\right)\left(1-\psi-\VP~,-n^i+\VV^i+\phi~n^i\right)
\equiv\left(1+\dnu~,-n^i-\dea^i\right)~,
\eneq
where we defined the perturbations $(\dnu,~\dea^i)$ in the photon wavevector.
In a homogeneous universe, the photon energy is redshifted 
$E=\hbar\omega\propto1/a$ 
as the Universe expands, and hence the constant factor $(\NC\omega a)$
can be effectively removed by choosing $\NC=1/\overline{(\omega a)}$. 
The choice of the
normalization in an inhomogeneous universe may be made only at one spacetime
point, as the combination $(\omega a)$ is {\it not} constant any more
\cite{YOZA14}, and this residual perturbation affects the parametrization
$(\dnu,~\dea^i)$ of the photon wavevector at the perturbation 
level.\footnote{Depending on the choice of the normalization condition,
the expressions of~$\dnu$ and~$\dea^i$ differ at each spacetime point,
causing some confusion and difficulty in comparing results. 
However, only the background~$n^i$ will be used for the luminosity distance 
calculations at the linear order in perturbations.}
Since the degree of freedom in the conformally transformed wavevector has
no physical significance, the final expression of observable quantities
is independent of the normalization choice. However, this choice 
$\overline{(\NC\omega a)}=1$ is the
most convenient and widely adopted in literature, while the choice differs
at the perturbation level.
We leave the normalization choice unspecified at the perturbation level
and show that the final results are indeed independent of the normalization
condition.

Given the observer four velocity and the photon wavevector, we can define
the observed photon vector in a FRW frame as
\beeq
\NN^a\equiv{k^a\over k^b\UU_b}+\UU^a=\frac1a\left(\VP~,~n^i-\phi~n^i\right)
~,\qquad \NN^a\NN_a=1~,\qquad \NN^a\UU_a=0~,
\eneq
and this vector becomes in the rest frame of the observer:
\beeq
N_L^a={k_L^a\over (k\cdot u)_L}+u_L^a=\left(0~,~\nhat\right)~,
\eneq
properly describing the observed direction of the photon path. However,
the observed photon vector~$N^a$ is written in a FRW frame, and its spatial
component is different from the observed angle~$\nhat$ in the observer
rest frame. Furthermore,
we can parallelly transport $N^a$ along the photon path
to describe the ``observed'' photon direction at each spacetime point
of the photon path. With respect to the conformally transformed metric,
the observed photon vector is 
\beeq
\CNN^a=a\NN^a={\CK^a\over\CK^b\CU_b}+\CU^a=
\left(\VP~,~n^i-\phi~n^i\right)~,\qquad \CNN_a=(-\VP~,~n_i+\phi~n_i)~,
\eneq
independent of the normalization~$\NC$.

\subsection{Deviations in the Photon Path}
\label{ssec:dist}
Given the expression of the photon wavevector, we want to integrate it 
over the affine parameter to
obtain the coordinate expression for the light propagation and relate it
to the observable quantities such as the observed redshift and the luminosity
distance. Eventually, this would provide connections to the different
approaches to modeling the luminosity distance.
The light propagates in a homogeneous universe without any distortion in path,
and the integration of the photon wavevector over the affine
parameter yields the propagation path at a given affine parameter:
\beeq
\bar x^a_\cc=\left(\bar\eta_\cc~,~\bar x^i_\cc\right)=\bar x^a_o+
\int_0^\cc d\cc'
~\hat{\bar k}^a_{\cc'}=\left(\bar\eta_o+\cc,~-\cc~n^i\right)~,
\eneq
where we set zero the affine parameter at the observer $\cc_o=0$. This
background relation in turn defines the affine parameter in relation to
the conformal time and the comoving distance as
\beeq
\cc=\bar\eta_\cc-\bar\eta_o=-\rbar_\cc~,\qquad
\rbar_z=\int_0^z{dz'\over H(z')}~,\qquad 1+z={1\over a(\bar\eta_z)}~,
\eneq
where the redshift parameter~$z$ is related to the affine parameter via
the conformal time $\cc_z=\bar\eta_z-\bar\eta_o$. We used a bar to 
represent quantities evaluated at~$\cc$ in a homogeneous universe and those
in an inhomogeneous universe will be represented without a bar.

The deviation from the straight path can be computed in a similar manner
by integrating the perturbations $(\dnu,\dea^i)$ 
in the photon wavevector over the affine parameter. These perturbations
obey the geodesic equations, and the temporal and spatial components
of the geodesic equations are
\beeq
0=\CK^a\CK^0_{\;\; ;a}={d\over d\cc}\dnu+\GG^0~,\qquad\qquad
0=\CK^b\CK^i_{\;\; ;b}=\left(-n^{i\prime}
 + n^jn^i_{\;\;|j}\right)-{d\over d\cc}\dea^i+\GG^i~,
\eneq
where the background relation of the spatial component defines the
photon path in a homogeneous universe as
\beeq
0={d\over d\cc}=\hat{\bar k}^a{\partial\over\partial x^a}=
{\partial \over \partial\eta}-n^j{\partial\over\partial x^j}~,
\eneq
and we defined two metric perturbations in the geodesic equations
\cite{YOZA14,YOO14a}
\bear
\GG^0&\equiv&\hat\Gamma^0_{ab}\CK^a\CK^b=
\psi'-2\psi_{,i}n^i+\phi'=2{d\over d\cc}\psi-(\psi-\phi)'~,\\
\GG^i&\equiv&\delta\left(\hat\Gamma^i_{ab}\CK^a\CK^b\right)=
\psi^{,i}- 2\phi'n^i+2\phi_{,j}n^jn^i-\phi^{,i}
=\left(\psi-\phi\right)^{,i}-2n^i{d\over d\cc}\phi~.
\enar 
With the conformal transformation, the background relation of the geodesic
equations is automatically satisfied. The perturbations to the photon
wavevector can be formally integrated over the affine parameter as
\bear
\label{dnuint}
\dnu_\cc-\dnu_o&=&-\int_0^\cc d\cc'~\GG^0
=2\psi_o-2\psi_\cc-\int_0^{\rbar_\cc} 
d\rbar~\left(\psi-\phi\right)'~,\\
\dea^i_\cc-\dea^i_o&=&\int_0^\cc d\cc'~\GG^i
=-2n^i\left(\phi_\cc-\phi_o\right)
-\int_0^{\rbar_\cc}d\rbar~\left(\psi-\phi\right)^{,i}~,
\enar
where we replaced the integration over the affine parameter~$\cc$ with the
integration over the comoving distance~$\rbar$, 
representing the background photon path. However,
it is noted that the integration over the affine parameter ($d\cc$)
represents the evaluation of the integrands along the photon path $x^a_\cc$, 
not along the background path $\bar x^a_\cc$,
although we will only need to consider the background path, as the
integrands are already at the linear order in perturbations.

One further integration of the perturbations in the photon wavevector
yields the deviation of the photon path from the background relation:
\bear
\delta x^a_\cc-\delta x^a_o&=&\left(\dT_\cc-\dT_o~,~\delta x^i_\cc\right)
=\left(\int_0^\cc d\cc'~\dnu~,~-\int_0^\cc d\cc'~\dea^i\right)~,\\
\label{eq:dtdt}
\delta\eta_\cc-\dT_o&=&-\rbar_\cc\left(2\psi+\dnu\right)_o+\int_0^{\rbar_\cc}
d\rbar\bigg[2\psi+(\rbar_\cc-\rbar)(\psi-\phi)'\bigg]~,\\
\delta x^i_\cc&=&\rbar_\cc\left(\dea^i+2\phi~ n^i\right)_o
-\int_0^{\rbar_\cc}d\rbar\left[2\phi~ n^i+(\rbar_\cc-\rbar)\left(\psi-\phi
\right)^{,i}\right] ~,
\enar
where the spatial position at the observer can always be set zero ($x^i_o=0$)
due to
symmetry and the conformal time at the observer in an inhomogeneous universe
deviates from its background value~$\bar\eta_o$ by
\beeq
\label{eq:dTdT}
\dT_o=-\int_0^{\bar\eta_o}d\eta\left(a\psi\right)=\int_0^\infty dz~
{\psi_o(z) \over H(z)}~,
\qquad\qquad \bar\eta_0=\int_0^\infty {dz\over H(z)}~.
\eneq
The light propagation is now expressed in terms of the metric perturbations
and the perturbations $(\dnu,\dea^i)$
to the photon wavevector at the observer position (or at some spacetime point).
Since the photon wavevector follows the null geodesic, these perturbations
are subject to the null condition:
\beeq
0=\CK^a\CK_a=2\left(n^i\dea_i-\dnu-\psi+\phi\right)~,
\eneq
and the remaining degrees of freedom in these perturbations $(\dnu,\dea^i)$
are eliminated
by the normalization condition in the conformal transformation in 
Eq.~\eqref{normal}. Therefore, the photon path is completely specified given
the metric perturbations and is independent
of our parametrization of the photon wavevector.

Given the observed photon direction~$\nhat$, 
it proves convenient to decompose the
deviation of the photon path into one along the line-of-sight direction and
one perpendicular to it:
\bear
\drr_\cc&\equiv&n_i\delta x^i_\cc=-\int_0^\cc d\cc'n_i\dea^i
=\dT_o-\dT_\cc+\int_0^{\rbar_\cc}d\rbar~(\psi-\phi)~,\\
\delta x^\perp_\cc
&=&\rbar_\cc(\dtt_\cc,~\sin\theta~\dpp_\cc)\equiv\delta x^i_\cc
{\nP}_i=\rbar_\cc\dea^i_o{\nP}_i
-\int_0^{\rbar_\cc}d\rbar~\left({\rbar_\cc-\rbar\over\rbar}\right)
\hat\nabla(\psi-\phi)~,
\enar
where the line-of-sight direction is $\nhat=(\theta,\phi)$ in spherical
coordinates, $\nP^i$ is a direction perpendicular to~$\nhat$,
the angular gradient is~$\hat\nabla$,
and the perpendicular component of $\delta x^i_\cc$ 
is also characterized
in spherical coordinates by using $(\dtt,\dpp)_\cc$.
These two components vanish at the observer position $\cc=0$.
Given the normalization condition in Eq.~\eqref{normal}, the first component
in~$\delta x^\perp_\cc$ is
\beeq
\label{kori}
\rbar_\cc\dea^i_o{\nP}_i=-\rbar_\cc \VV^i_o{\nP}_i~,
\eneq
independent of the normalization condition at the perturbation order.

\subsection{Observed Redshift and Geometric Distortions}
\label{ssec:redshift}
An important observable in large-scale structure probes is the redshift
of luminous sources such as galaxies. The photon wavelength is stretched
due to the cosmic expansion, as it propagates throughout the Universe, and
the observed redshift parameter is simply a measure of how much it has been
stretched from its emission in the source rest frame
along its journey to reach the
observer:
\beeq
1+\zz\equiv{\lambda_o\over\lambda_s}={\omega_s\over\omega_o}
={\left(k^a\UU_a\right)_s\over\left(k^a\UU_a\right)_o}~,
\eneq
where we used that the angular frequency~$\omega$ measured in the
rest frames of the source and the observer is a Lorentz scalar
\beeq
\omega=-\eta_{ab}u_L^a k^b_L=-g_{ab}u^ak^b~.
\eneq
Using the conformally transformed wavevector, we can define a useful 
quantity $\dhnu$ that vanishes in the background in relation to the 
normalization condition in Eq.~\eqref{normal}:
\beeq
-\CK^a\CU_a=-\NC a(k^a\UU_a)=\NC a\omega\equiv1+\dhnu~,\qquad\qquad
\dhnu\equiv\dnu+\psi+\VP=n_i\dea^i+\phi+\VP~,
\eneq
and the observed redshift is then
\beeq
1+z={a_o\over a_s}{(\CK^a\CU_a)_s\over(\CK^a\CU_a)_o}=
{a_o\over a_s}{1+\dhnu_s\over1+\dhnu_o}~.
\eneq
It is apparent that the observed redshift is affected not only by the cosmic
expansion, but also by the line-of-sight velocity and the gravitational
redshift, in the presence of inhomogeneities. To separate the background
and the perturbation quantities, we define the distortion $\dz$ in the redshift
as
\beeq
\label{dzdef}
1+z\equiv{1+\dz\over a_s}~,\qquad \qquad
\dz=\HH_o\dT_o+\dhnu_s-\dhnu_o~,
\eneq
where we noted that $\eta_o=\bar\eta_o+\dT_o$ and $a(\bar\eta_o)=1$.
Using the relation for~$\dnu$ in Eq.~\eqref{dnuint}, the distortion in the
redshift can be readily computed as \cite{SAWO67}
\beeq
\dz=\HH_o\dT_o+\left(\psi-\VP\right)_o-\left(\psi-\VP\right)_z
-\int_0^{\rbar_z} d\rbar~(\psi-\phi)'~,
\eneq
where we replaced the source position denoted by the subscript~$s$ with
the redshift parameter at the linear order in perturbations. It is noted that
the expression is independent of our parametrization of the photon wavevector
and the conformal transformation.

In the cosmological context, the redshift parameter is the only physically
meaningful way to characterize its distance from the observer. Other parameters
such as the affine parameter or the coordinate positions are both 
gauge-dependent and unobservable quantities, inadequate to describe physical
quantities. To accommodate this point in our previous calculations, 
we define a series of perturbation
quantities with respect to the background quantities evaluated at the
observed redshift. In particular, they concern with the relation of the
apparent and real positions of the source, representing the geometric
distortions in an inhomogeneous universe:
First, the affine parameter at the source position
is split into~$\cc_z$ corresponding to its observed redshift and the residual 
perturbation~$\Dcc_s$:
\beeq
\cc_s\equiv\cc_z+\Dcc_s~,\qquad \qquad
\cc_z=\bar\eta_z-\bar\eta_o~,\qquad\qquad 1+z={1\over a(\bar\eta_z)}~.
\eneq
Second, the time coordinate~$\eta_s$ of the source is also split into
one $\bar\eta_z$ associated with the redshift 
and the residual perturbation~$\DT_z$:
\beeq
\eta_s=\bar\eta_s+\dT_s\equiv \bar\eta_z+\DT_z~,\qquad\qquad
\DT_z=\bar\eta_s-\bar\eta_z+\dT_s=\Dcc_s+\dT_s~.
\eneq
Given the definition of the distortion~$\dz$ of the redshift, the deviation
$\DT_z$ in the time coordinate is further
related to the distortion in the redshift as
\beeq
\DT_z={\dz\over\HH}~.
\eneq
Third, we might define the deviation $\DX^i_z$ in the spatial coordinates
of the source, 
in a similar way $\DT_z$ is defined, but $\DX^i_z$ will not appear in any
of our equations at the linear order in perturbations.
Finally, the geometric
deviations $(\drr_\cc,\delta x^\perp_\cc)$ of the photon path
also need to be expressed with respect to the background position 
$\bar x^i_z=\rbar_z n^i$ at the observed redshift:
\beeq
\drr_z\equiv n_ix^i_{s}-\rbar_z=\drr_{s}-\Dcc_s
=\dT_o-{\dz\over\HH}+\int_0^{\rbar_z}d\rbar~(\psi-\phi)~,\qquad\qquad
\delta x^\perp_z=\delta x^\perp_{s}~,
\eneq
where the deviation perpendicular to the line-of-sight is unaffected
as there is no transverse component in the background.
At the observer position~$\cc=0$, these perturbation quantities become
\beeq
\up{at}~\cc=0~(\zz=0)~:~~ \dz=\HH_0\dT_o~,
\quad \Dcc_s=0~,\quad \DT_z=\dT_z=\dT_o~,\quad \drr_z=\delta x^\perp_z=0~.
\eneq

\subsection{Gravitational Lensing Magnification}
\label{ssec:kappa}
Gravitational lensing describes the angular distortion in the photon path.
The photon path is affected by the matter
fluctuation and the metric perturbations along its path, and the observed
angle of the source is non-trivially related to the position we would measure
in the absence of such perturbations (see, e.g., \cite{BLNA86,BASC01}). 
However, at the linear order in 
perturbations, the deviation from its unperturbed path is rather simple and
well studied in literature. 
In view of our geometric description of the light propagation, 
gravitational lensing deals with the angular
distortion $(\dtt,\dpp)$ without concerning the radial distortion~$\drr$
in our formalism. In particular, only the lensing magnification
will be needed in our application to the luminosity distance.

In general, the angular distortion in gravitational lensing is described
by the inverse magnification matrix, or the deformation matrix~$\DD$,
providing the relation between the observed angular position $(\ttt,\pp)$
at the observer to the angular position $(\ttt+\dtt,\pp+\dpp)$ of the
source. The deformation matrix is conventionally decomposed as
\beeq
\DD\equiv
{\partial(\ttt+\dtt,\pp+\dpp)\over \partial(\ttt,\pp)}
\equiv\left(\begin{array}{cc}1&0\\0&1\end{array}\right)
-\left(\begin{array}{cc}\kappa+\gamma_1&w+\gamma_2\\-w+\gamma_2&
\kappa-\gamma_1\end{array}\right)~,
\eneq
where the deviation from the identity captures the angular distortion
in gravitation lensing with the convergence~$\kappa$, the rotation~$w$, 
and the shear~$(\gamma_1,\gamma_2)$. Since the surface brightness is
conserved, the geometric enhancement of
the solid angle $d\Omega_s$ of the source results in the gravitational lensing
magnification, and it is the inverse Jacobian of the deformation matrix:
\beeq
{d\Omega_o\over d\Omega_s}={1\over \up{det}~\DD}={1\over (1-\kappa)^2-
\gamma^2+w^2}\simeq1+2~\kappa+\OO(2)~.
\eneq
At the linear order in perturbation, the lensing magnification is proportional
to the convergence. Given the definition of the deformation matrix, its
determinant can be computed as \cite{YOZA14}
\beeq
\up{det}~\DD
={\sin(\ttt+\dtt)\over \sin\ttt}
\left[1+{\partial\over\partial\ttt}\dtt+{\partial\over\partial\pp}\dpp
+{\partial\over\partial\ttt}\dtt{\partial\over\partial\pp}\dpp
-{\partial\over\partial\ttt}\dpp{\partial\over\partial\pp}\dtt
\right]~,
\eneq
valid to {\it all orders} in perturbations, when the angular 
distortion $(\dtt,\dpp)$ is also defined non-perturbatively.
The lensing convergence at the linear order is, therefore, derived as
\beeq
\kappa=-{1\over2}\left[\left(\cot\ttt+{\partial\over\partial\ttt}
\right)\dtt+{\partial\over\partial\pp}\dpp\right]
=-{\VP}_o+\int_0^{\rbar_z}d\rbar\left({\rbar_z-\rbar\over
\rbar_z\rbar}\right)\hat\nabla^2\left({\psi-\phi\over2}\right)~,
\eneq
where the line-of-sight velocity at the observer position arises
from Eq.~\eqref{kori} and the angular Laplacian is
\beeq
\hat\nabla^2=\left(\cot\ttt+{\partial\over\partial\ttt}
\right){\partial\over\partial\ttt}
+{1\over\sin^2\ttt}{\partial^2\over\partial\pp^2}~.
\eneq
The line-of-sight velocity at the observer position is often ignored
in the gravitational lensing convergence. Consequently, it is missing 
in the luminosity distance calculation as well 
\cite{BEGAET12a,BEMAET12a,BEGAET13}. In \cite{BEDUET14}, all the observer 
velocity contributions (in $\dz$, $\drr$, and~$\kappa$) are ignored by hand
(see \cite{KAHU15a} for interesting discussion of the velocity contribution
to the luminosity distance).

\section{Geometric Approach to the Luminosity Distance}
\label{sec:geom}
The apparent source position $\bar x_s^a$
is inferred by using the observable quantities such as
the observed redshift and the observed angular position. In 
Sec.~\ref{sec:formalism}, the distortion in the apparent source position
compared to the real source position is geometrically decomposed with
the radial distortion~$\drr$ and the angular distortion~$\kappa$,
accounting for the
distortion~$\dz$ in the observed redshift as the ``distortion''
in the time direction.
Since gravity achromatically affects the light propagation, our geometric
approach should provide a good physical description of cosmological probes 
including the luminosity distance fluctuations.

The crucial element in the geometric approach is the relation of the source
position to the physical quantities. In \cite{YOFIZA09,YOO10}, the covariant
expression was first applied in cosmology to compute the physical volume
occupied by the source~$x^a_s$ with its inferred position~$\bar x^a_s$.
Such covariant expression is essential to ensuring the gauge-invariance of 
the expression of a physical volume in terms of the observable quantities
\cite{YOO10}. Adopting this approach, \citet{JESCHI12} first presented
a covariant expression of the physical area, occupied by the source~$x^a_s$
with its inferred position~$\bar x^a_s$, and the area is further defined
as one perpendicular to the photon propagation. This altogether provides
a key element for computing the luminosity distance in the geometric approach.

\subsection{Covariant Expression}
\label{ssec:cov}
The luminosity distance is related to the angular
diameter distance by the reciprocity relation
\beeq
\label{dAdL}
\dA(z)={\dL(z)\over(1+z)^2}~,
\eneq
at the given observed redshift, and this relation is exact to all orders
in perturbations. Therefore, the fluctuation in the angular diameter distance
is equivalent to that in the luminosity distance:
\beeq
1+\ddL={\dL(z)\over\bar D_L(z)}={\dA(z)\over\bar D_A(z)}=1+\delta\mathcal{D}_A
~,
\eneq
providing a simple way to derive the luminosity distance using our description
of the geometric distortion.
The angular diameter distance is the distance, at which a unit area $\dAp$
at the observed redshift~$z$ is subtended by a solid angle $d\Omega_o$
at the observer:
\beeq
\label{defdA}
\dAp=\dA^2(z)d\Omega_o~.
\eneq
Since the area~$\dAp$ in the source rest frame is perpendicular to the
four velocity~$u^a$ at the source and also to the observed photon 
vector~$\NN^a$
transported to the source position, the area spanned by the observed
angle can be computed in a covariant way \cite{JESCHI12,YOZA14,YOO14a} as
\beeq
\label{covE}
\dAp=\sqrt{-g}~\varepsilon_{dabc}\UU^d\NN^a
{\partial x^b\over\partial\ttt}{\partial x^c\over\partial\pp}~d\ttt~d\pp ~,
\eneq
where $\varepsilon_{abcd}$ 
is the Levi-Civita symbol with $\varepsilon_{0123}=1$. Using the relations
\beeq
\label{eq:useful}
a_s={1+\dz\over1+z}~,\qquad \sqrt{-g}=a^4_s(1+\dg)~,\qquad \dg=\psi+3\phi~,
\qquad \bar D_A(z)={\rbar_z\over1+z}~,
\eneq
we can re-arrange the the covariant equation by
\beeq
\dA^2(\zz)=\bar D_A^2(\zz){(1+\dg)(1+\dz)^2\over\rbar_z^2\sin\ttt}
\left(\varepsilon_{dabc}\CU^d\CNN^a
{\partial x^b\over\partial\ttt}{\partial x^c\over\partial\pp}\right)~,
\eneq
and the round bracket can be further expanded in its Levi-Civita indices as
\beeq
\left(\varepsilon_{dabc}\CU^d\CNN^a
{\partial x^b\over\partial\ttt}{\partial x^c\over\partial\pp}\right)=
\varepsilon_{0i j k }\CNN^i
{\partial x^j \over\partial\ttt}{\partial x^k \over\partial\pp}+
\varepsilon_{0i j k }(-\psi)\CNN^i
{\partial x^j \over\partial\ttt}{\partial x^k \over\partial\pp}+
\varepsilon_{kabc}\VV^k\CNN^a
{\partial x^b\over\partial\ttt}{\partial x^c\over\partial\pp}~,
\eneq
where the third component is already at the second order in perturbations.
Since the second component is already at the linear order, we need to
evaluate the remaining term at the background level:
\beeq
\varepsilon_{0i j k }(-\psi)\hat{\bar\NN}^i
{\partial\bar x^j \over\partial\ttt}{\partial\bar x^k \over\partial\pp}=
-\psi~\nhat\cdot\left(\rbar_z{\partial\over\partial\ttt}\nhat\times \rbar_z
{\partial\over\partial\pp}\nhat\right)=-\psi~\rbar_z^2\sin\ttt~.
\eneq
Similarly, the first component can be computed by decomposing it according to
the perturbation orders as
\bear
&&\varepsilon_{0i j k }\CNN^i
{\partial x^j \over\partial\ttt}{\partial x^k \over\partial\pp}=
\rbar_z^2\sin\ttt+
\varepsilon_{0i j k }\CNN^{i(1)}
{\partial \bar x^j \over\partial\ttt}{\partial \bar x^k \over\partial\pp}+
\varepsilon_{0i j k }\hat{\bar\NN}^i\left(
{\partial \delta x^j \over\partial\ttt}{\partial\bar x^k \over\partial\pp}+
{\partial \bar x^j \over\partial\ttt}{\partial\delta x^k\over\partial\pp}
\right)~ \\
&&\qquad=\rbar_z^2\sin\ttt~\left(1-\phi\right)
+\rbar_z\nhat\cdot\left(
{\partial \over\partial\ttt}\delta\bdi{x}\times
{\partial \over\partial\pp}\nhat+
{\partial\over\partial\ttt}\nhat\times{\partial\over\partial\pp}\delta\bdi{x}
\right)
=\rbar_z^2\sin\ttt~\left(1-\phi+2~{\drr_z\over\rbar_z}
-2~\kappa\right)~,\nonumber
\enar
where the vector product is essentially the distortion in the geometric
volume. Therefore, the fluctuation in the angular diameter distance amounts to
\beeq
\delta\mathcal{D}_A=
\frac12\dg+\dz+\frac12\left(-\psi-\phi+2~{\drr_z\over\rbar_z}-2~\kappa\right)
=\phi+\dz+{\drr_z\over\rbar_z}-\kappa~,
\eneq
and it is equivalent to the fluctuation in the luminosity distance
$\delta\mathcal{D}_A=\ddL$. The fluctuations arise because of the (comoving)
volume distortions decomposed into the radial component$~\drr$ and the angular
component~$\kappa$, the relation to the proper volume by the metric 
determinant~$\dg$, and finally
the use of the observed redshift~$\dz$ in the luminosity distance.
We will use our geometric approach to connect different methods for
computing the luminosity distance in the following sections.

\subsection{Standard Ruler}
Another approach to computing the physical area was developed 
in \cite{SCJE12a} under the name of ``cosmic ruler.'' The idea is to relate
the (known) scale of a standard ruler placed at source~$x^a_s$ to the inferred 
position~$\bar x^a_s$ in terms of the geometric distortions.
Similar in spirit to obtaining the covariant expression in Eq.~\eqref{covE},
 this 
method computes the length of a standard ruler with two end points of the
ruler described by the observed angle $(\ttt,\pp)$ and $(\ttt+d\ttt,\pp+d\pp)$.
The original method \cite{SCJE12a}
computes the distortion in three-dimensional space, but
here we will focus on one that is relevant to computing the luminosity 
distance.

Assuming two ends points $x^a_1$ and $x^a_2$ of a standard ruler
are so close to each other 
that they both are at the same observed redshift~$z$, the 
difference~$\delta\bar x^a$
in the inferred positions can be computed as
\beeq
\bar x_1^a=[\bar\eta_z,~\rbar_z n^i]~,\qquad
\bar x_2^a=[\bar\eta_z,~\rbar_z (n^i+\Delta n^i)]~,\qquad
\delta\bar x^a\equiv\bar x^a_1-\bar x^a_2=(0,-\rbar_z\Delta n^i)~,
\eneq
where the difference in the observed angle
$\Delta n^i=(d\ttt,d\pp)$ in a spherical coordinate will be
taken to be zero. Since two apparent angular directions are unit vectors,
we have $\delta\bar x^a\propto\Delta n^i\perp n^i$. Given the apparent
positions of the standard ruler, we can derive the relation of its apparent
scale~$R_z$ to the known scale of the ruler~$R$:
\bear
\label{eq:ruler}
R^2&=&\PP_{ab}(x_1-x_2)^a(x_1-x_2)^b=
\PP_{ab}\left(\delta\bar x^a\delta\bar x^b+2\delta\bar x^a
\DX^b+\DX^a\DX^b\right)~\nnn
&=&R^2_z(1+2~\dz)+2a^2(\bar\eta_z)
\left(\phi~\gbar_{ij}\delta\bar x^i\delta\bar x^j
+\gbar_{ij}\delta\bar x^i\DX^j\right)~,
\enar
where we ignored the second order terms,
the apparent scale of the ruler is related to the angular diameter
distance in a homogeneous universe
\beeq
R_z^2\equiv a^2(\bar\eta_z)\gbar_{ij}\delta\bar x^i\delta\bar x^j
=\bar D_A^2(z)\gbar_{ij}\Delta n^i\Delta n^j~,
\eneq
the difference $\DX^a$ in the real and the apparent positions of two
end points are
\beeq
x_1^a=\bar x_1^a+\DX^a_1~,\qquad x_2^a=\bar x_2^a+\DX^a_2~,\qquad
\DX^a\equiv \DX^a_1-\DX^a_2~,
\eneq
and the projection tensor is
\beeq
\PP_{ab}=g_{ab}+u_au_b~,\qquad\qquad \PP^a_a=3~.
\eneq
Two end points are projected to a hypersurface orthogonal to~$u^a$, because the
standard ruler is defined in a local rest frame of an observer~$u^a$.
With the approximation that the scale of the ruler is small, 
the projection tensor $\PP_{ab}$ at~$x^a_1$ was used without ambiguity.
Similarly, no ambiguity arises in the line-of-sight direction ($=\nhat_1$).

It is apparent in Eq.~\eqref{eq:ruler} that the scale of the ruler 
is set by its apparent angular scale $R^2\propto\Delta n^2$, except the last
term with~$\DX^i$. Since $\DX^i$ vanishes as the angular
scale $\Delta n^i=(d\ttt,d\pp)$ is taken to be zero, the leading
contribution in proportion to~$\Delta n^i$ can be computed as
\beeq
\DX^i\simeq\delta \bar x^j{\partial\over\partial x^j}\DX^i\simeq
-\rbar_z\Delta n^j
{\partial\over\partial x^j}\DX^i\simeq
-\rbar_z\Delta n^j{\partial\over\partial x^j}\left(n^i\drr+\delta 
x^\perp_z\right)+\mathcal{O}(\Delta n^2)~,
\eneq
and the last term in Eq.~\eqref{eq:ruler} then becomes
\beeq
2a^2(\bar\eta_z)\gbar_{ij}\delta\bar x^i\Delta x^j=\bar D_A^2(z)
\left[2~{\drr\over\rbar_z}\gbar_{ij}\Delta n^i\Delta n^j+
\gbar_{ij}\Delta n^i\Delta n^k\hat\nabla_k(\dtt,~\sin\ttt~\dpp)^j\right]~,
\eneq
where we took the limit $x^a_2\RA x^a_1$ and the component is expressed 
in spherical coordinates.
Therefore, putting it altogether, the relation in Eq.~\eqref{eq:ruler} 
can be used to compute a unit physical area~$\dAp$ spanned by the angular
scale $(d\ttt,d\pp)$ as
\beeq
\label{ruleout}
\dAp=\bar D_A^2(z)\left(1+2~\dz+2~\phi+2~{\drr_z\over\rbar_z}-2~\kappa
\right)d\Omega=\dA^2(z)~d\Omega~,
\eneq
consistent with the result in previous section.

The covariant expression in Sec.~\ref{ssec:cov}
directly computes a unit physical
area $\dAp$ in an observer rest frame~$u^a$, 
corresponding to the angular scale~$(d\ttt,d\pp)$. The standard ruler
calculation in \cite{SCJE12a} starts with a scale in an observer rest frame,
demanding that it be observed at the angular scale~$(d\ttt,d\pp)$. The 
resulting expressions are naturally described by the geometric distortions
in both cases.\footnote{\citet{SCJE12a} further considers a case in which 
the scale
of a standard ruler evolves in time, i.e., by replacing~$\dz$ in
Eq.~\eqref{ruleout} with
\beeq
\dz\left(1-{\partial\ln R^2\over\partial \ln a}\right)~.
\eneq
However, such term would break the
gauge-invariance of the luminosity distance calculation, unless the frame
of such evolution is specified, e.g., the proper-time hypersurface of an
observer:
\beeq
\dz-\dz_p\left({\partial\ln R^2\over\partial \ln a}\right)~,
\eneq
where the subscript indicates that a gauge choice is made, while~$\dz$
without it is left unspecified as other terms in Eq.~\eqref{ruleout}.
This gauge issue is also resolved in their later work \cite{JESC13}.
}

\section{Sachs Approach to the Luminosity Distance: Optical Scalar Equation}
\label{sachs}
In this section, we present the optical scalar equation and its relation to
the luminosity distance. This approach was first taken by 
Sasaki \cite{SASAK87} and later extended by Umeh et~al.
\cite{UMCLMA14a,UMCLMA14b}.

\subsection{Optical Scalar Equation}
The light propagation in an inhomogeneous universe was presented in 
Sec.~\ref{sec:formalism}. Now we consider a bundle of light rays 
and how its properties like the expansion~$\theta$ and the shear~$\sigma$ 
propagate along the geodesic (see \cite{EHLER61,ELLIS71}
for review). In the rest frame
of an observer described by the four velocity~$\UU^a$, a unit area~$\dAp$
can be uniquely defined as one perpendicular to the photon propagation~$\NN^a$,
and the projection into this two-dimensional hypersurface is described by
the projection tensor:
\beeq
\HH_{ab} \equiv g_{ab} + \UU_a \UU_b - \NN_a \NN_b = g_{ab} +
\frac{k_a k_b}{(k\cdot \UU)^2} - 2 \frac{\UU_{(a}k_{b)}}{k\cdot \UU}~,
\eneq
where the two-dimensional projection tensor is orthogonal to the observer and
the light propagation direction
\beeq
0=\HH_{ab}\UU^b=\HH_{ab}\NN^b=\HH_{ab}k^b~,\qquad\qquad\HH^a_a=2~.
\eneq
As the distortion tensor $k_{a;b}$ is three-dimensional, we project the
distortion tensor into the two-dimensional hypersurface and decompose it
in terms of the expansion~$\theta$ and the projected shear~$\sigma_{ab}$ as
\beeq
\HH_{ac}\HH_{bd}k^{c;d}={1\over2}\theta~\HH_{ab}+\sigma_{ab}~,
\eneq
where the expansion and the amplitude of the shear are independent of
its projection tensor
\beeq
\theta=\HH^a_c\HH^d_a{k^c}_{;d}={k^a}_{;a}~,\qquad\qquad
2\sigma^2\equiv\sigma_{ab}\sigma^{ab}=k_{a;b}k^{a;b}-\frac12\theta^2~.
\eneq
When the photon wavevector is parametrized to be orthogonal to the 
hypersurface of constant phase~$\vartheta$, as in Eq.~\eqref{wavephy}, 
there is no asymmetric part of the distortion tensor (or rotation).
The shear vanishes in the background.

Parametrizing the photon wavevector $k^a=dx^a/d\oo$ with (physical) affine 
parameter~$\oo$, 
the propagation of the distortion tensor can be readily computed as
\beeq
\label{eq:propagation}
{D\over d\oo}\left({k^a}_{;b}\right)=
k^c{k^a}_{;bc}=-k_{c;b}k^{a;c}-R^a_{dbc}k^ck^d~,
\eneq
where we used the geodesic equation and the definition of the Riemann tensor.
Contracting the two indices, we can derive the propagation equation for the
expansion (or {\it Sachs equation} \cite{SACHS61,KRSA66})
\beeq
{d\over d\oo}\theta+\frac12\theta^2+2\sigma^2+R_{ab}k^ak^b=0~,
\eneq
where it is noted that
the propagation equation and the expansion itself are independent
of a projection tensor~$\HH_{ab}$.

In particular, the expansion of light rays represents the change of 
a unit area~$\dAp$, swept by the bundle of neighboring
light rays, and hence it is 
related to the angular diameter distance as in Eq.~\eqref{defdA}:
\beeq
\label{ttaa}
\theta={d\over d\oo}\ln\dAp={d\over d\oo}\ln \dA^2~.
\eneq
A formal integration of the expansion over the affine parameter yields
the angular diameter distance
\beeq
\dA(\oo)=\dA(\epsilon)\exp\left[\frac12\int_\epsilon^\oo d\oo'~
\theta(x^a_{\oo'})\right]~,
\eneq
where the affine parameter $\oo=\epsilon$ will be eventually taken to be zero,
representing the observer position.
The luminosity distance can be readily obtained by
using the reciprocity relation in Eq.~\eqref{dAdL}.
As we have noted in Sec.~\ref{sec:formalism}, the calculations are greatly
simplified, when we work in the conformally transformed metric, as it is
devoid of the scale factor and its derivatives like the Hubble parameter.
We will do so in deriving the above solution for the angular diameter distance.

\subsection{Conformal Transformation and Angular Diameter
Distance}
\label{ssec:conf}
Given the conformal transformation of the metric in Eq.~\eqref{metric},
we have proved in Sec.~\ref{sec:formalism}
that some quantities between these two metrics
are trivially related by the scale factor~$a$, while some quantities are 
{\it not}: 
\beeq
g_{ab}=a^2\hat g_{ab}~,\qquad \UU^a=a\CU^a~,\qquad \NN^a=a\CNN^a~,\qquad
\CK^a=\NC a^2k^a~.
\eneq
With the above relations, the projection tensor transforms trivially as 
$\HH_{ab}=a^2\hat\HH_{ab}$. 
We need to exercise caution in deriving the governing equations
and the relevant quantities in the conformally transformed metric.
For example, the expansion~$\theta$ is transformed differently, 
as the covariant derivatives in two metrics are non-trivially related.

Noting that the metric tensors are parallelly transported in each metric
\beeq
0=\nabla_ag_{bc}=\hat\nabla_a\hat g_{bc}~,
\eneq
we can show that
the covariant derivative $\hat\nabla_a$ in the conformally transformed metric,
when acted on a four vector~$\CK^a$,
is related to the covariant derivative~$\nabla_a$ in the original metric as
\beeq
\label{confrel}
\hat\nabla_a\CK^b=\nabla_a\CK^b+C^b_{ac}\CK^c~,\qquad\qquad
C^c_{ab}=\HH\left(g_{ab}g^{c0}-\delta^0_a\delta^c_b-\delta^c_a\delta^0_b
\right)~,
\eneq
where $C^b_{ac}$ is the symmetric tensor, connecting two derivative operators.
Contracting two indices, the expansion~$\CT$ in the conformally transformed
metric can be related to the expansion~$\theta$ as
\beeq
\CT=\hat\nabla_a\CK^a=
\nabla_a\left(\NC a^2k^a\right)-4\HH\CK^0=\NC a^2\theta-2\HH\CK^0
={d\oo\over d\cc}~\theta-{d\over d\cc}\ln a^2~.
\eneq
Using the relation of the expansion to a unit area in Eq.~\eqref{ttaa},
we derive its counterpart in the conformally transformed metric
\beeq
\label{eq:CTDA}
\CT={d\over d\cc}\ln\left({\dAp\over a^2}\right)
={d\over d\cc}\ln\left({\dA\over a}\right)^2~,
\eneq
and therefore, the angular diameter distance can be obtained by using~$\CT$
as
\beeq
\label{eq:angfor}
\dA(\cc)={a(\cc)\over a(\epsilon)}~
\dA(\epsilon)\exp\left[\frac12\int_\epsilon^\cc d\cc'~\CT(x^a_{\cc'})\right]~.
\eneq

The propagation of the expansion~$\CT$ should be governed by one similar
to the Sachs equation. Indeed,
the derivation of the Sachs equation is generic, such that it only depends
on the parametrization of the photon wavevector, the metric tensor and its
derivatives. Therefore, it is obvious that the Sachs equation takes the
same form in the conformally transformed metric:
\beeq
\label{eq:Sachs}
{d\over d\cc}\CT+\frac12\CT^2+2\hat\sigma^2+\XX=0~,
\qquad\qquad \XX\equiv\hat R_{ab}\CK^a\CK^b~,
\eneq
where the expansion and the shear are defined in the same way, but in terms
of the conformally transformed wavevector~$\CK^a$.

\subsection{Sasaki 1987 \cite{SASAK87}}
\label{sasaki}
In the pioneering paper \cite{SASAK87}, the complete derivation of the
luminosity distance was presented with the general metric representation,
and the gauge invariance of its expression was shown by explicitly transforming
each component under a general coordinate transformation. Despite its
perfection, the work \cite{SASAK87} remains as one of the most 
incomprehensible in the luminosity distance literature.

Here we reproduce the derivation of the luminosity distance in \cite{SASAK87},
paying particular attention to its connection to our geometric approach.
Given the solution in Eq.~\eqref{eq:angfor}, we need to compute the integration
of the expansion and the angular diameter distance $\dA(\epsilon)$ near
the observer position.
First, consider a bundle of light rays from a distant source converging at the
observer position, where the expansion of light rays becomes infinite.
The expansion at the background can be readily obtained by integrating
the Sachs equation with the boundary condition as
\beeq
{d\hat{\bar\theta}\over d\lambda}=-\frac12\hat{\bar\theta}^2~,\qquad\qquad
\hat{\bar\theta}=\frac2\cc~,
\eneq
where both $\hat\sigma$ and $\hat R_{ab}\CK^a\CK^b$ in the Sachs equation
vanish at the background. The integration in the angular diameter
distance in Eq.~\eqref{eq:angfor} trivially becomes
\beeq
\exp\left[\frac12\int_\epsilon^\cc d\cc'~\hat{\bar\theta}\right]
={\cc\over \epsilon}~,
\eneq
where $\cc=\epsilon$ will be taken to be zero.

Next, consider a local Lorentz frame of the observer to compute the relation of
the angular diameter distance $\dA(\epsilon)$ near the observer to 
the photon path. In the observer's rest frame, the angular 
diameter distance at $\cc=\epsilon$ is essentially the physical distance 
$\Delta r$ that light travels for an infinitesimal amount of time $\Delta t$, 
corresponding to the affine parameter $\cc=\epsilon$:
\beeq
\label{dtdtdt}
\omega=-\eta_{ab}k^a_Lu^b_L=k_L^0={dt\over d\cc}{1\over\NC a^2}~,\qquad
\Delta t=(\NC a^2\omega)_o\epsilon~,\qquad \dA(\epsilon)=|\Delta t|
=-(\NC a^2\omega)_o\epsilon~.
\eneq
Using this relation and expanding the integration in Eq.~\eqref{eq:angfor} 
to the linear order in perturbation, the angular diameter distance at the
source $\cc_s$ can be obtained as
\bear
\dA&=&-a_s(\NC a\omega)_o\cc_s
\left(1+\frac12\int_0^{\cc_s}d\cc~\delta\CT\right)
=-\left({1+\dz\over1+z}\right)
\left(1+\dhnu_o\right)\left(\cc_z+\Dcc_s\right)
\left(1+\frac12\II\right) \nnn
&=&{\rbar_z\over1+z}\left(1+\dz+\dhnu_0+{\Dcc_s\over\cc_z}+\frac12\II\right)
=\bar D_A(1+\delta\mathcal{D}_A)~,
\enar
where the infinitesimal affine parameter~$\epsilon$ cancels,
we replaced each term $a_s$, $(\NC a\omega)_o$,
and $\cc_s$, and we defined the integration of
the expansion perturbation~$\delta\CT$ 
\beeq
\II\equiv \int_0^{\cc_s}d\cc~\delta\CT(x^a_\cc)~.
\eneq

The expansion perturbation~$\delta\CT$
is subject to the linear-order Sachs equation and its solution given
the background solution~$\hat{\bar\theta}$ is
\beeq
{d\over d\cc}\delta\CT+\hat{\bar\theta}~\delta\CT+\delta\XX=0~,
\qquad\qquad
\delta\CT_\cc=-{1\over\cc^2}\int_0^\cc d\cc'(\cc')^2
\delta\XX_{\cc'}~.
\eneq
As the Ricci scalar $\hat R=\hat R^a_a$
in the transformed metric vanishes in the
background, the integrand for the expansion perturbation is
\bear
\label{eq:JR}
\delta\XX_\cc&=&\XX(x^a_\cc)=
\Delta\left(\psi-\phi\right)-\left(\psi-\phi\right)_{,i|j}n^in^j
-2{d^2\over d\cc^2}\phi \nnn
&=&{2\over\rbar}(\psi-\phi)'-{2\over\rbar}{d\over d\cc}(\psi-\phi)
+{1\over\rbar^2}\hat\nabla^2(\psi-\phi)
-2{d^2\over d\cc^2}\phi~. 
\enar
Using the formal solution of~$\delta\CT_\cc$, the integration of
the expansion perturbation can be computed as
\beeq
\label{eq:III}
\II=\int_0^{\cc_s}d\cc~\delta\CT_\cc=-\int_0^{\cc_s}d\cc\left({\cc_s
-\cc\over\cc_s\cc}\right)\cc^2\delta\XX_\cc
\equiv \II_A+\II_B+\II_C+\II_D~,
\eneq
where we split the integration into four components, corresponding to
each component in the integrand in Eq.~\eqref{eq:JR}
\bear
\II_A&=&-{2\over\rbar_z}\int_0^{\rbar_z}d\rbar~(\rbar_z-\rbar)
\left(\psi-\phi\right)'~,\qquad\qquad
\II_B=2(\psi-\phi)_o-{2\over\rbar_z}\int_0^{\rbar_z}d\rbar~(\psi-\phi)~,\\
\II_C&=&-\int_0^{\rbar_z}d\rbar\left({\rbar_z-\rbar\over\rbar_z\rbar}\right)
\hat\nabla^2(\psi-\phi)=-2\kappa-2{\VP}_o~,\qquad\qquad
\II_D=2(\phi+\phi_o)-{4\over\rbar_z}\int_0^{\rbar_z} d\rbar~\phi~.\nonumber
\enar
So, the integration~$\II$ concerns the angular distortion~$\kappa$ in
addition to several contributions of the gravitational potential.
Using the relation for $\dT_z-\dT_o$ in Eq.~\eqref{eq:dtdt}, the residual
perturbation~$\Dcc_s$ of the source~$\cc_s$ is related to the radial
distortion~$\drr_z$ as
\bear
-\Dcc_s&=&\dT_z-{\dz\over\HH}=\drr_z-\rbar_z(2\psi+\dnu)_o+\int_0^{\rbar_z}
d\rbar\left[(\psi+\phi)+(\rbar_z-\rbar)(\psi-\phi)'\right]~,\\
{\Dcc_s\over\cc_z}&=&{\drr_z\over\rbar_z}-\dhnu_o+{\VP}_o+\phi-\frac12
\left(\II_A+\II_B+\II_D\right)~.
\enar
Finally, putting it all together, we derive
\beeq
\label{sasakiresult}
\delta\mathcal{D}=\dz+\dhnu_o+{\Dcc_s\over\cc_z}+\frac12\II
=\dz+{\drr_z\over\rbar_z}-\kappa+\phi~,
\eneq
and the result is consistent with the previous calculations in
our geometric approach. 

In \cite{SASAK87}, the normalization convention
was taken as $\dhnu_s=0$ at the source position, but as we demonstrated,
the final expression is independent of the normalization convention.
After the complete derivation of the expression with the general metric,
additional computation was performed \cite{SASAK87} in the conformal
Newtonian gauge, in which the coordinate lapse~$\dT_o$ at the observer
position was removed by taking the angle average of~$\ddL$ and demanding
it should vanish at $z=0$:
\beeq
0\equiv\lim_{z\RA0}\int{d\Omega\over4\pi}~\ddL(z)={3\over\eta_0}\dT_0+\psi_o
+{\eta^2_o\over4}\left(\psi_{,ij}n^in^j\right)_o~,\qquad
\dT_o\equiv-{\eta_o\over3}\left[\psi+{\eta^2\over18}\Delta\psi\right]_o~.
\eneq
This is in conflict with Eq.~\eqref{eq:dTdT}. 
The coordinate lapse at the observer position 
is important in ensuring the gauge-invariance of the full expression~$\ddL$,
and it cannot be removed this way.

\subsection{Umeh, Clarkson, Maartens 2014 \cite{UMCLMA14a}}
\label{lumdist:umeh}
Despite the difference in the master equations,
the approach taken in \cite{UMCLMA14a,UMCLMA14b} is indeed identical to
one in \cite{SASAK87}, all of which are based on the Sachs 
equation~\eqref{eq:Sachs}. Therefore, this method can reproduce the
correct luminosity distance, while a few terms are neglected in
\cite{UMCLMA14a,UMCLMA14b}.

Given the relation of the expansion~$\CT$ to the angular diameter
distance in Eq.~\eqref{eq:CTDA}, we take the derivative 
with respect to the affine parameter:
\beeq
{d\over d\cc}\CT={d^2\over d\cc^2}\ln\left({\dA\over a}\right)^2=
2~{a\over\dA}{d^2\over d\cc^2}\left({\dA\over a}\right)
-\frac12\CT^2~,
\eneq
and using the Sachs equation to remove the expansion~$\CT$, 
we derive the differential equation for the angular diameter distance:
\beeq
{d^2\over d\cc^2}\left({\dA\over a}\right)=-\frac12~\XX
\left({\dA\over a}\right)~,
\eneq
where we kept the terms up to the linear order in perturbations.
Since the source term in the right-hand side vanishes in the background,
we can readily derive the background solution for the differential equation:
\beeq
\bar{\mathcal{D}}_A=-a\cc=a\rbar_\cc~.
\eneq
With the background solution in the source term, we can integrate 
the equation twice to obtain the linear-order solution as
\bear
\left({d\over d\cc}{\dA\over a}\right)_\cc^{(1)}
&=&\left({d\over d\cc}{\dA\over a}\right)_o^{(1)}
+\frac12\int_0^\cc d\cc'~\cc'~\XX_{\cc'}
~,\\
\left({\dA\over a}\right)_s^{(1)}&=&\cc_s\left({d\over d\cc}{\dA\over a}
\right)_o^{(1)}+\frac12\int_0^{\cc_s}d\cc~(\cc_s-\cc)\cc~\XX_\cc
=\cc_s\left[\left({d\over d\cc}{\dA\over a}
\right)_o-\frac12\II\right]~,~~~~~
\enar
where we used $\dA(\cc_o)=0$ and the definition of~$\II$ in Eq.~\eqref{eq:III}.
Therefore, the angular diameter distance is
\beeq
\dA=-a_s\cc_s\left[1-\left({d\over d\cc}{\dA\over a}\right)_o^{(1)}
+\frac12\II\right]=\bar D_A(z)\left[1+\dz+{\Dcc_s\over\cc_z}
-\left({d\over d\cc}{\dA\over a}\right)_o^{(1)}+\frac12\II\right]~,
\eneq
where we expressed the scale factor~$a_s$ and the affine parameter~$\cc_s$
of the source in terms of its observed redshift and residual perturbations.
Comparing to the result in Eq.~\eqref{sasakiresult}, it is clear that
the boundary condition for the derivative term at the observer should be
\beeq
\left({d\over d\cc}{\dA\over a}\right)_o^{(1)}=-\dhnu_o~,
\eneq
and the solution would be consistent with our previous calculations.

According to \cite{KRSA66} (their equation~34), the derivative of the angular
diameter distance at the observer is related to the photon energy as
\beeq
\left({d\over d\oo}\dA\right)_o=-\omega_o~.
\eneq
The derivative with respect to the conformally transformed affine parameter
yields
\beeq
{d\over d\cc}\left({\dA\over a}\right)=\NC a{d\over d\oo}\dA-H\dA~,\qquad\qquad
\left({d\over d\cc}{\dA\over a}\right)_o=-\left(\NC a\omega\right)_o=-\left(
1+\dhnu\right)_o~,
\eneq
and indeed the boundary condition at the observer is consistent with
our expectation.

Despite the derivation here, there exist a few inconsistencies in the results
of \cite{UMCLMA14b} --- they have ignored several terms at the observer
position:
\beeq
\dT_o=0~,\qquad \dnu_o=0~,\qquad \dea^i_o=0~.
\eneq
The first term $\dT_o$ in Eq.~\eqref{eq:dTdT} cannot be set zero in
the conformal Newtonian gauge. Either the second term or the third term
may be set zero by using the normalization condition~$\NC$, but not both
at the same time. The absence of these terms 
breaks the gauge invariance of the luminosity distance.
Finally, equation~(101) in \cite{UMCLMA14b} 
\beeq
{1\over \bar D_A}\left({\dA\over a}\right)^{(1)}_s=-\psi+{\VP}_o
+{1\over\rbar_z}\int_0^{\rbar_z}d\rbar~\left[2\psi-\left({\rbar_z-\rbar
\over\rbar^2}\right)\hat\nabla^2\psi\right]~,
\eneq
should correspond in our notation to
\beeq
{1\over \bar D_A}\left({\dA\over a}\right)^{(1)}_s=\dhnu_o+\frac12\II 
=2\psi_o-\psi+{\VP}_o-{1\over\rbar_z}\int_0^{\rbar_z}d\rbar
\left[\left({\rbar_z-\rbar\over\rbar}\right)\hat\nabla^2\psi
+\left(\rbar_z-\rbar\right)2\psi'\right] ~,
\eneq
where we took $\dnu_o=0$ and $\phi=-\psi$ to facilitate the comparison,
as adopted in \cite{UMCLMA14b}.

\section{Jacobi Mapping Approach to the Luminosity Distance}
\label{Jacobi}
Similar in spirit to the Sachs approach based on the optical scalar equation,
another approach to modeling the luminosity distance is to utilize the
Jacobi mapping, or the {\it geodesic deviation} equation 
(see, e.g., \cite{SESCEH94,LECH06} for reviews). We first present
the propagation equations for the Jacobi field and then derive the luminosity
distance by using the Jacobi field.

\subsection{Geodesic Deviation Equation and Jacobi Matrix}
\label{ssec:dev}
We derive how two neighboring light rays from the same source
propagate and how they are related to the boundary condition. 
Consider two light rays at the same affine parameter~$\oo$, separated by
an infinitesimal distance $\delta x^a(\oo)=\xi^a\delta\Sigma$, where 
the connecting vector~$\xi^a$ is called a Jacobi field and $\Sigma$
parametrizes the separation~$\delta x^a$ of two rays along the same affine
parameter~$\oo$. We will derive the Jacobi field along the main ray,
working in the limit $\delta\Sigma\rightarrow0$. Since the Jacobi field
connects two rays at the same affine parameter or the same phase~$\vartheta$
of the wave propagation, it is orthogonal to the photon wavevector, and
the Jacobi field can be written in general as
\beeq
\label{JACOBF}
\xi^a k_a=0~,\qquad\qquad \xi^a\equiv \xi^I[e_I]^a+\xi^0k^a~,\qquad\qquad
\HH^{ab}=[e_I]^a[e_J]^b\delta^{IJ}~,
\eneq
where we introduced orthonormal tetrads $[e_I]^a$ with $I=1,2$ that specify
the two-dimensional hypersurface (characterized by the projection
tensor $\HH_{ab}$) orthogonal to the photon propagation direction.
Four tetrads (one time-like $[e_t]^a=\UU^a$ and three space-like $[e_i]^a$)
in Section~\ref{sec:formalism} form a local orthonormal basis in the rest
frame of the observer described by~$\UU^a$. Here two space-like
tetrads $[e_I]^a$ are constructed among the three space-like tetrads,
but orthogonal to the photon propagation direction~$\NN^a$ in the rest frame.
Furthermore, while the nonvanishing component~$\xi^0$ is consistent with
the orthogonality condition of the Jacobi field, it bears no physical 
relevance to quantities of
our interest, and we set $\xi^0=0$.\footnote{This is in fact 
possible by constructing a projected Jacobi field $\xi_\perp^a=\HH^a_b\xi^b$.
However, 
since we will exclusively work on the projected field, we simply set $\xi^0=0$
and call $\xi^a$ a (projected) Jacobi field.}

The propagation equation for the Jacobi field (or the geodesic deviation
equation) can be derived in a way similar to the derivation of
Eq.~\eqref{eq:propagation} (e.g., \cite{SESCEH94,LECH06}):
\beeq
\label{eq:deviation}
{D\over d\oo}\xi^a=k^b\nabla_b\xi^a=\xi^b\nabla_bk^a~,\qquad\qquad
{D^2\over d\oo^2}\xi^a=k^c\nabla_c(k^b\nabla_b\xi^a)=-R^a_{bcd}k^b\xi^ck^d~,
\eneq
where we have used the symmetry of~$\xi^a$ along the geodesic and
the geodesic equation for~$k^a$. By transporting the local tetrads along
the geodesic, the propagation equations can be converted into the those
for the components of the Jacobi field:
\beeq
{d\over d\oo}\xi^I=\JB^I_J\xi^J~,\qquad \qquad \qquad
{d^2\over d\oo^2}\xi^I=-\JR^I_J\xi^J~,
\eneq
where we have defined the projected tensors
\beeq
\JB^I_J=(\nabla_bk^a)[e^I]_a[e_J]^b~,\qquad\qquad
\JR^I_J=(R^a_{bcd}k^bk^d)[e^I]_a[e_J]^c~.
\eneq
In relation to the propagation equation, the Jacobi matrix~$\JD$ 
simply connects the Jacobi field at one position to another position at~$\oo$: 
\beeq
\xi^I(\oo) \equiv \JD^I_J(\oo)\dot\xi^J_o~,\qquad \qquad
\dot\xi^I_o\equiv{d\over d\oo}\xi^I\bigg|_{\oo_o}~,
\eneq
where we set the boundary condition at the observer, as the light rays
converge at the observer position $\xi^I(\oo_o)=0$.
In terms of the Jacobi matrix, the propagation equations can be readily
expressed as
\beeq
{d\over d\oo}\JD^I_J=\JB^I_K\JD^K_J~,\qquad\qquad
{d^2\over d\oo^2}\JD^I_J = -\JR^I_K\JD^K_J~,
\eneq
where the boundary condition of the Jacobi field translates into
the boundary condition for the Jacobi matrix:
\beeq
\JD^I_J(\oo_o)=0~,\qquad \qquad
{d\over d\oo}\JD^I_J\bigg|_{\oo_o}=\delta^I_J~.
\eneq
As we derived in Sec.~\ref{sasaki},
the non-vanishing boundary condition $\dot\xi^I_o$ can be obtained by
considering the Jacobi field near the observer position $\oo=\epsilon$,
in which 
\beeq
\xi^I(\epsilon)=|\Delta t|n^I~,\qquad\qquad \Delta t=\omega_o\epsilon~,
\qquad\qquad
\dot\xi^I_o=-\omega_o n^I~,
\eneq
where $\Delta t$ is again the infinitesimal time (or distance with $c=1$)
corresponding to the propagation of light from the origin to the physical
affine parameter $\oo=\epsilon$
and $n^I=(\theta,\phi)$ is the observed angle in a spherical coordinate.
Since the physical area at~$\oo$ is simply 
\beeq
\label{eq:JACOBL1}
\dAp(\oo)=\xi^1\xi^2\bigg|_\oo=\det~\JD(\oo)~\dot\xi^1_o\dot\xi^2_o~,
\eneq
we related the angular diameter distance to the Jacobi map as
\beeq
\label{eq:JACOBL2}
\dA^2(\oo)=\det~\JD(\oo)\omega_o^2~.
\eneq
\bigskip

\subsection{Jacobi Matrix in a Conformal Transformed Metric}
\label{ssec:jaconf}
As evident in Sec.~\ref{ssec:conf}, 
the conformal transformation in Eq.~\eqref{metric} gives rise to non-trivial
relations to the derivative operators and its associated products. Furthermore,
a geodesic path is {\it no} longer geodesic once conformally transformed,
except when it is a null geodesic. Consequently, the propagation equations
for the Jacobi matrix need to be carefully transformed, according to the
change in the derivative operators.

In a similar way the four velocity $\UU^a$ transforms, we
define conformally transformed tetrads $[\htet_I]^a$ as
\beeq
[\htet_t]^a=\CU^a=a\UU^a~,\qquad\qquad [\htet_I]^a\equiv a[e_I]^a~,
\eneq
and concordantly the Jacobi field in this basis is
\beeq
\xi^a=\xi^I[e_I]^a=\left({\xi^I\over a}\right)[\htet_I]^a\equiv
\hxi^I[\htet_I]^a~.
\eneq
While there is {\it no} unique definition for the conformally transformed
Jacobi field (e.g., $\hat \xi^a$), our definition of $\hat \xi^I=\xi^I/a$
is the only choice, with which the propagation equations take the same form
in the conformally transformed metric. We will refer~$\hxi^I$ to the
conformally transformed Jacobi field.

First, we consider the propagation of
the Jacobi field with respect to the conformally transformed affine parameter:
\beeq
{D\over d\cc}\xi^a={d\oo\over d\cc}{D\over d\oo}\xi^a=
\NC a^2~\xi^b\nabla_b k^a
=\xi^b\hat\nabla_b\CK^a-\HH\xi^0\CK^a+\xi^a{d\over d\cc}\ln a~,
\eneq
where we used the relation in Eq.~\eqref{confrel} and the derivative of
the scale factor is
\beeq
{d\over d\cc}a={da\over d\eta}{d\eta\over d\cc}=a'\CK^0=a\HH \CK^0~.
\eneq
Absorbing the derivative term to the LHS, 
the propagation of the Jacobi field in the
conformally transformed metric is then
\beeq
{D\over d\cc}\left({\xi^a\over a}\right)=\left({\xi^b\over a}\right)
\hat\nabla_b\CK^a-\HH\left({\xi^0\over a}\right)\CK^a~,
\eneq
and the presence of the second term in contrast to Eq.~\eqref{eq:deviation}
highlights the difference associated with the conformal transformation.
Now, we decompose the Jacobi field in terms of the conformally transformed
tetrads $[\htet_I]^a$:
\beeq
\up{LHS}={[\hat e_I]^a\over a}{d\over d\cc}
\hat \xi^I+{\hat\xi^I\over a}\HH\CK^0[\hat e_I]^a-{\hat\xi^I[\hat e_I]^a
\over a}{d\over d\cc}\ln a~,\qquad
\up{RHS}=\left({\hat\xi^I[\hat e_I]^b\over a}
\right)\hat\nabla_b\CK^a-\HH\left({\hat\xi^I[\hat e_I]^0\over a}\right)\CK^a~,
\eneq
where the conformally transformed tetrads obey
\beeq
{D\over d\oo}[e_I]^a={D\over d\oo}[e_I]_a=0~,\qquad
{D\over d\cc}[\hat e_I]^a=\HH\CK^0[\hat e_I]^a~,\qquad
{D\over d\cc}[\hat e_I]_a=-\HH\CK^0[\hat e_I]_a~.
\eneq
Note that the conformally transformed tetrads are {\it no} longer parallelly
transported along the photon path.
Multiplying the tetrads on both sides and renaming the indicies, we arrive
at
\beeq
{d\over d\cc}\hat\xi^I=(\hat\nabla_b\CK^a)[\hat e_J]^b[\hat e^I]_a\hat\xi^J
\equiv\hat\JB^I_J\hxi^J~,
\eneq
where the projected tensor~$\hat\JB^I_J$ takes the same form as in the
original definition, but with the conformally transformed metric.
This justifies our choice of the definition~$\hxi^I$. 
Taking another derivative, we 
complete the derivation of the remaining propagation equation:
\bear
{d^2\over d\cc^2}\hxi^I&=&
-\CK^a{}_{;c}\CK^c{}_{;b}[\htet_J]^b[\htet^I]_a\hxi^J+
\hat R^a_{dcb}\CK^c\CK^d[\htet_J]^b[\htet^I]_a\hxi^J+(\hat\nabla_b\CK^a)
\hat\JB^I_J{d\over d\cc}\hxi^J~\nnn
&=&-\hat R^a_{dbc}\CK^c\CK^d[\htet_J]^b[\htet^I]_a\hxi^J
=-\hat\JR^I_J\hxi^J~,
\enar
where the derivative terms of the tetrads are cancelled.

In relation to the propagation equations in the conformally transformed
metric, the conformally transformed Jacobi matrix~$\hat\JD$ 
satisfies the same form of the propagation equations
\beeq
\hat\xi^I(\cc)=\hat\JD^I_J(\cc)\dot{\hat\xi}_o^J~,\qquad\qquad
{d\over d\cc}\hat\JD^I_J=\hat\JB^I_K\hat\JD^K_J~,\qquad\qquad
{d^2\over d\cc^2}\hat\JD^I_J = -\hat\JR^I_K\hat\JD^K_J~,
\eneq
with the same boundary conditions for $\hat\JD^I_J$. 
However, the boundary condition for
the conformally transformed Jacobi field $\hat\xi^I$ is different:
\beeq
\hxi^I(\cc_o)=0~,\qquad \qquad
\dot\hxi^I_o\equiv{d\over d\cc}\hxi^I\bigg|_{\cc_o}=-\left(\NC a\omega\right)_o
n^I=-(1+\dhnu_o)n^I~,
\eneq
where the normalization in the background is set
$\overline{\mathbb{C}a\omega}=1$. Therefore, the angular diameter distance
is then
\beeq
\label{eq:JACOBL}
\dA^2(\cc)=a_\cc^2\det~\hat\JD(\cc)\left(\NC a\omega\right)_o^2
=a_\cc^2(1+\dhnu_o)^2\det~\hat\JD(\cc)~.
\eneq
\bigskip

\subsection{Fluctuation in the Luminosity Distance}
The fluctuation in the luminosity distance will be computed by using the
relation of the Jacobi map to the angular diameter distance. To compute
the Jacobi map, the propagation equations need to be solved given the 
boundary condition. The source term of the propagation equation is
\beeq
\hat\JR^I_J=(\hat R^a_{bcd}\CK^b\CK^d)[\htet^I]_a[\htet_J]^c
=\frac12\delta^I_J\hat\JR+(\hat C^a_{bcd}\CK^b\CK^d)[\htet^I]_a[\htet_J]^c~,
\eneq
where we defined the conformal (Weyl) tensor
\beeq
\hat C^a_{bcd} =\hat R^a_{bcd}-\frac12\left(\delta^a_c\hat R_{bd}
+\hat g_{bd}\hat R^a_c-\hat g_{bc}\hat R^a_d-\delta^a_d \hat R_{bc} \right)
+\frac{\hat R}6 \left(\delta^a_c\hat g_{bd}-\delta^a_d\hat g_{bc} \right)~,
\eneq
and the trace of the source tensor is 
$\hat\JR^I_I\equiv\hat\JR=\hat R_{ab}\CK^a\CK^b$ as in Eq.~\eqref{eq:JR}.
With the vanishing source tensor in the background, the propagation equation
can be trivially integrated to yield
\beeq
{d\over d\cc}\hat{\bar\JD}^I_J(\cc)=\delta^I_J~,\qquad\qquad
\hat{\bar\JD}^I_J(\cc)=\cc~\delta^I_J=-\rbar_\cc\delta^I_J~.
\eneq
At the linear order in perturbation, the source tensor
contributes to the propagation equations, and the integration over the 
background solution gives
\beeq
{d\over d\cc}\hat\JD^I_J{}^{(1)}(\cc)=-\int_0^\cc d\cc'~\cc'~
\hat\JR^I_J(\cc')~,\qquad \qquad
\hat\JD^I_J{}^{(1)}(\cc_s)=-
\int_0^{\cc_s}d\cc~(\cc_s-\cc)\cc~\hat\JR^I_J(\cc)~.
\eneq
The determinant of the Jacobi map is therefore 
\beeq
\det\hat\JD(\cc_s)=\cc_s^2+\cc_s\hat\JD^I_I(\cc_s)=\cc_s^2\left[1-
\int_0^{\cc_s}d\cc\left({\cc_s-\cc\over\cc_s\cc}\right)\cc^2
~\hat\JR(\cc)\right]
=\cc_s^2\left(1+\II\right)~,
\eneq
where the integral is exactly the integration of the expansion 
perturbation~$\delta\CT$ in Eq.~(\ref{eq:III}). Putting it altogether
and keeping the linear order terms, the angular diameter distance is
derived as
\beeq
\dA=a_s(1+\dhnu_o)~\det{}^{1/2}\hat\JD(\cc_s)=-a_s\cc_s\left(1+\dhnu_o
+\frac12\II\right)=\bar D_A(z)\left(1+\dz+{\drr_z\over\rbar_z}-\kappa+\phi
\right)~,
\eneq
consistent with the previous derivations.

\subsection{Bonvin, Durrer, Gasparini 2006 \cite{BODUGA06}}
The Jacobi mapping approach was first introduced in
modeling the luminosity distance \cite{BODUGA06}. This approach provides
a physically simple description of the light propagation measured by
an observer along the photon path, and it can be readily generalized to
the weak lensing formalism \cite{SESCEH94,LECH06,BEBOET12}.
The Jacobi field in \cite{BODUGA06} is computed by using the Jacobi (four)
vector~$\xi^a$ in Eq.~\eqref{JACOBF}, rather than its projected 
component~$\xi^I$ in our approach. Consequently, the Jacobi matrix
becomes a 4-by-4 matrix, in which only two-dimensional subspace carries
the relevant information. In \cite{BODUGA06}, the determinant $|\hat J|$ 
of the Jacobi matrix in the conformally transformed metric corresponds
to something similar, but different from the determinant of our
Jacobi matrix:
\beeq
|\hat J|^{1/2}=-\cc_s(1+\dhnu_s)\left(1-2\psi+\Delta\alpha\right)
=-\cc_s\left(1+\dhnu_s+\frac12\II\right)\neq\det\hat\JD^{1/2}~,
\eneq
where $\Delta\alpha$ is the notation used in \cite{BODUGA06}
and their $\hat w=1+\dhnu$. However, when the physical angular diameter
distance is computed as
\beeq
\mathcal{D}_A=a_s~{\hat w_o\over\hat w_s}|\hat J|^{1/2}
=-a_s\cc_s\left(1+\dhnu_o+\frac12\II\right)=
\bar D_A(z)\left(1+\dz+{\drr_z\over\rbar_z}-\kappa+\phi
\right)~,
\eneq
it can reproduce the correct expression consistent with our previous
results, if the terms~$\dT_o$ neglected throughout their calculations
are reinstated. As their main interest lies in computing the angular
power spectrum, the absence of the coordinate lapse~$\dT_o$ in their expression
of the luminosity distance affects nothing; the monopole and the dipole
are often {\it not} part of the power spectrum analysis.

\section{Geodesic Light Cone Approach to the Luminosity Distance}
\label{GLC}
The geodesic light cone (GLC) approach to modeling the luminosity distance
and other observables in cosmology was first introduced in \cite{GAMAET11},
and it was further extended \cite{BEGAET12a,BEMAET12,BEGAET13,BEGAET13a}
to higher-order calculations of the luminosity distance.

The geodesic light cone coordinates is similar in spirit to our geometric
approach, 
in which the building blocks of theoretical descriptions are the basic 
observable quantities such as
the observed redshift, the observed angular position of sources, the
observed flux, and so on. It differs, however, in that the GLC
approach incorporates 
this idea in its coordinate system $x^a= (\ww,\tau,\tilde\theta^I)$, 
where $\ww$ describes the phase of past 
light cones ($\ww\sim\vartheta$ in Eq.~\eqref{phase}),  
$\tau$ is the proper time of observers moving with
time-like velocity~$\UU^a$, and $\TT^I$ with $I=1,2$ describes the observed
direction of the light propagation ($\TT^I\sim n^I$). The FRW metric
in an inhomogeneous universe is then described by a GLC coordinate as
\beeq
ds^2=\Upsilon^2d\ww^2-2\Upsilon d\ww d\tau + \gamma_{IJ}(d\TT^I-U^Id\ww)
(d\TT^J-U^Jd\ww)~,
\eneq
and the metric components in the background recover the usual FRW components
and justify their physical meaning
--- the proper time $\tau=t$, the scale factor $\Upsilon=a(\eta)$, the
phase of past light cones $\ww=\rbar+\bar\eta\propto\vartheta$, 
a spherical coordinate $\gamma_{IJ}d\TT^Id\TT^J=a^2\rbar^2d\Omega$ 
with the observed angles $\TT^I=n^I$, and the auxiliary function $U^I=0$.

In the presence of inhomogeneities in the Universe, their physical meaning is 
rather involved and needs clarification in relation to other gauge choice.
Transforming the conformal Newtonian gauge $y^a=(\eta,x,y,z)$
to the GLC coordinate $x^a=(\ww,\tau,\TT^I)$ at each spacetime point,
\beeq
g_\up{GLC}^{ab}(x^e)={\partial x^a\over\partial y^c}
{\partial x^b\over\partial y^d}~g^{cd}(y^e)=
-{1-2\psi\over a^2}{\partial x^a\over\partial\eta}
{\partial x^b\over\partial\eta}+{1-2\phi\over a^2}
\gbar^{ij}\partial_ix^a\partial_jx^b~,
\eneq
we derive a series of differential equations for the GLC metric components:
\bear
(\tau\tau)&:&~d\tau=a(1+\psi)d\eta~,\qquad\qquad\qquad
(ww):~{\partial\over\partial\eta_+}\bar w=1~,
\qquad{\partial\over\partial \eta_-}w=\frac12(\psi-\phi)~,\\
\label{gamma}
(wI)&:&~{\partial\over\partial\eta_-}\tilde\theta^I=\frac12\gbar^{IJ}
\partial_Jw~,\qquad\qquad (IJ):~
\gamma^{IJ}=\frac1{a^2}\left[\gbar^{IJ}(1-2\phi)
+\gbar^{IK}\partial_K\tilde\theta^J+\gbar^{KJ}\partial_K\tilde\theta^I
\right]~,~~~\\
(w\tau)&:&~
{1\over\Upsilon}=\frac1a\left(1-\psi+{\partial\over\partial \eta_+} w
+{\partial\over\partial \eta_-} w
-\frac1a{\partial\over\partial\rbar}\tau\right)
\equiv\frac1a(1-\delta\Upsilon)~,
\enar
where we defined the background lightcone variables
\beeq
{\partial \over \partial \eta_+}={1\over 2}
\left({\partial\over\partial\eta}+{\partial\over \partial\rbar}\right)~,
\qquad\qquad
{\partial \over \partial \eta_-}={1\over 2}
\left({\partial\over\partial\eta}-{\partial\over \partial\rbar}\right)
=\frac12{d\over d\lambda}=-\frac12{d\over d\bar r}~.
\eneq
Integrating these differential equations, we obtain
\bear
\tau &=& \int_0^\eta d\eta' a(1+\psi)~, \qquad\qquad
w =\eta_+-2\bar r\Psi_\up{av}~,\qquad\qquad
\Psi_\up{av}\equiv{1\over2\rbar}\int_0^{\bar r}d\bar r~(\psi-\phi)~,\\
\tilde\theta^I &=& \theta_o^I+\int_0^{\rbar}d\rbar'\gbar^{IJ}_{\rbar'}
\int_0^{\rbar'}d\rbar''\partial_J(\psi-\phi)~,\qquad\qquad
\delta\Upsilon=\psi+\int_0^{\rbar} d\rbar'(\psi-\phi)'-\VP~,
\enar
and the GLC variable~$U^I$ is {\it not} needed for our present purposes.

The advantage of the GLC coordinates is the simplicity in expression of
observable quantities, although they eventually need to be computed in any
of conventional choices of gauge conditions. The null vector~$k^a$ of a
past light cone is specified by the constant phase~$\vartheta\sim\ww$:
\beeq
k^a=g^{ab}\partial_b\ww=\left(0,-\frac1\Upsilon,0,0\right)~,
\qquad\qquad k_a=g_{ab}k^b=(1,0,0,0)~,
\eneq
where the components and indices are for GLC coordinates. Similarly, the
four velocity of geodesic flows with proper time~$\tau$ is 
\beeq
\UU_a={\partial_a\tau\over\sqrt{\partial\tau\cdot\partial\tau}}
=(0,-1,0,0)~,\qquad\qquad
\UU^a=g^{ab}\UU_b=\left(\frac1\Upsilon,~1,~{U^I\over\Upsilon}\right)~.
\eneq
Consequently, the observed redshift in GLC coordinates is
simply the ratio of~$\Upsilon$:
\beeq
1+z={(k_\mu u^\mu)_s\over(k_\mu u^\mu)_o}={\Upsilon_o\over\Upsilon_s}
={1\over a_s}\left(1+\HH_o\dT_o+\delta\Upsilon_o-\delta\Upsilon_s\right)~,
\eneq
and indeed its expression in terms of the conformal Newtonian gauge variables
matches the distortion in the observed redshift
\beeq
\HH_o\dT_o+\delta\Upsilon_o-\delta\Upsilon_s=\dz~,
\eneq
if the first term $\HH_o\dT_o$ neglected in \cite{BEMAET12,BEGAET13a}
is included.
Comparing Eq.~\eqref{dzdef}, it is evident that 
$(\delta\Upsilon_o-\delta\Upsilon_s)$ corresponds to
$(\dhnu_s-\dhnu_o)$ in our notation.

Finally, the fluctuation of the luminosity distance is again obtained
by computing the angular diameter distance. In GLC coordinates, a unit area
perpendicular to the light propagation in the rest frame of geodesic flows
is 
\beeq
\label{glcda}
\dAp=\dA^2d\Omega_o\propto \sqrt{|\gamma|}~d^2\TT~,\qquad
\qquad \gamma=\det~\gamma_{IJ}~,
\eneq
and the angular diameter distance is therefore
\beeq
\dA(\oo)=\dA(\epsilon)\left({|\gamma(\oo)|\over|\gamma(\epsilon)|}\right)
^{1/4}~,
\eneq
where we used the affine parameters to indicate where the GLC quantities
are evaluated.\footnote{In \cite{GAMAET11}, the angular diameter distance
is defined by lumping together
the quantities near the observer ($\oo=\epsilon$) as a proportionality
constant~$c$:
\beeq
\label{incnor}
\dA^2(\oo)\equiv c\sqrt{|\gamma(\oo)|}~,\qquad\qquad
c^{-1}=\sin\TT_o=\sin\theta_o~,
\eneq
and the constant was obtained by taking the limit $\oo\RA0$. However,
the computation of this limit is incorrect, because the observed photon
direction in FRW frame is different from that in the observer rest frame,
as shown in Eq.~\eqref{normal}. From Eq.~\eqref{ratio}, 
the correct constant is
\beeq
c={\left(1+\dhnu_o-\phi_o\right)^2\over\sin\theta_o}
={\left(1+{\VP}_o\right)^2\over\sin\theta_o}~.
\eneq
Compared to the derivation the luminosity distance in \cite{GAMAET11},
our derivation in Eq.~\eqref{glc} with the correct proportionality constant~$c$
yields additional terms $\dhnu_o-\phi_o={\VP}_o$ that cancel the extra
velocity at the observer position. This correct normalization is also obtained
in their later work \cite{FAGAET13}, but with different approach.}
Using Eq.~\eqref{gamma}, the angular determinant at a given spacetime point is
\beeq
\gamma^{-1}=\bar\gamma^{-1}\left[1-4\phi
+2\int_0^{\rbar}d\rbar'\left({\rbar-\rbar'\over\rbar\rbar'}\right)
\left({\partial^2\over\partial\theta^2}+{1\over\sin^2\theta}{\partial^2
\over\partial\phi^2}\right)(\psi-\phi)\right]\equiv{1-4\phi+4\tilde J_2\over
a^4r^4\sin^2\theta}~,
\eneq
and the ratio of two angular determinants is then
\beeq
\label{ratio}
{|\gamma(\oo)|\over|\gamma(\epsilon)|}
=\left({1-4\phi_o+0\over a_o^4r_o^4\sin^2\theta_o}\right)
\left({1-4\phi+4\tilde J_2\over a_s^4r_s^4\sin^2\theta_s}\right)^{-1} 
=\left({a_sr_s\over a_or_o}\right)^4(1+\phi-\phi_o-J_2)^4~,
\eneq
where we have defined $\tilde J_2$ and $J_2$ through the equations following
the convention in \cite{BEGAET12a}:
\beeq
J_2\equiv\tilde J_2+1-\sqrt{\sin\theta_s\over\sin\theta_o}=
\tilde J_2+\frac12\TT^{(1)}\cot\theta_o=\frac12
\int_0^{\bar r_s}d\bar r\left({\bar r_s-\bar r\over\bar r_s\bar r}
\right)\hat\nabla^2(\psi-\phi)=-\frac12\II_C~.
\eneq
Since the angular diameter distance $\dA(\epsilon)$ near the observer is
related to the distance $\Delta r=-(\NC a^2\omega)_o\cc_\epsilon$ 
that light travels for $\Delta t$ in Eq.~\eqref{dtdtdt},
the angular diameter distance can be readily computed as
\beeq
\mathcal{D}_A=-(\NC a^2\omega)_o\cc_\epsilon\left({a_sr_s\over a_or_o}\right)
\left(1+\phi-\phi_o+\frac12\II_C\right)
=a_sr_s\left(1+\dhnu_o\right)\left(1+\phi-\phi_o-\kappa-{\VP}_o\right)~,
\eneq
and the fluctuation in the luminosity distance is then
\beeq
\label{glc}
\ddL=\delta\dA=\dz+{\drr\over\rbar_z}+\phi-\kappa+
\left(\dhnu_o-\phi_o-{\VP}_o\right)~,
\eneq
where the comoving distance at origin is taken to be zero:
$r_o=-\cc_\epsilon\RA0$. 

Mind the presence of three extra terms in the parenthesis in Eq.~\eqref{glc},
compared to the previous 
derivations. In Eq.~\eqref{glcda}, we assumed that the GLC coordinates
$\TT^I$ are the observed angles~$n^I$, i.e., $d\Omega_o=d^2\TT$. However,
as apparent in Eq.~\eqref{normal}, the light propagation direction in 
GLC coordinates is indeed proportional to
\beeq
\TT^I\propto n^I+\dea^I~, 
\eneq
and therefore, the equality $d\Omega_o=d^2\TT$ holds only if the
photon normalization is set at the observer position
\beeq
0=\dea^i_o=n^i(\dhnu-\phi)_o-\VV^i_o~.
\eneq
This condition yields
\beeq
\dhnu_o=\phi_o+{\VP}_o~,\qquad\qquad \dnu_o=-(\psi-\phi)_o~,
\eneq
putting the derivation in GLC coordinates consistent with other approaches.

The GLC attempts to utilize the observed angle in its angular coordinate
to describe observable quantities, but the original calculations in 
\cite{GAMAET11,BEGAET12a,BEMAET12a} neglected the subtle difference 
in the observed angle in the observer rest frame and that in the FRW frame,
in addition to the absence of the coordinate lapse term~$\dT_o$.
However, the normalization condition for the angular variables 
was fixed in \cite{FAGAET13},
which brings the expression fully consistent with other approaches.
Our normalization condition provides another way to derive the correct
expression for the luminosity distance.

\section{Discussion}
\label{sec:discussion}
We have computed the luminosity distance by adopting four different approaches
in literature (the geometric, the Sachs, the Jacobi mapping, and the geodesic 
light cone approaches) 
and presented a unified treatment of the luminosity distance calculation,
facilitating the comparison of the different approaches
and verifying the sanity of each approach to modeling the luminosity distance.
The advantage and disadvantage of each approach is as follows. We started
with the geometric approach in \cite{JESCHI12,YOO14a,YOZA14} as a base for our
calculations. With the equivalence
principle, gravity affects everything in the same way,
including all spectrum of the 
electromagnetic waves (hence achromatic), and 
the fluctuation in the luminosity distance should therefore be associated 
with the geometric distortions of the photon path and its flux. In this regard,
the geometric approach 
provides the simplest description and the physically intuitive interpretation 
of the
luminosity distance, in which the change in the luminosity distance arises
from the volume distortion (the radial~$\drr$ and the angular~$\kappa$) 
in conjunction with the change~$\dz$ in the observed redshift.

The Sachs approach is somewhat mysterious in its original form \cite{SASAK87},
in deriving the angular diameter distance in Eq.~\eqref{eq:angfor}
and the angular diameter distance $\dA(\epsilon)$ at the origin,
which we clarify in relation to the geometric approach. With the clarification
in this work,
the rules of computing the luminosity distance are straightforward in this
approach, and the seemingly different approach taken in 
\cite{UMCLMA14a,UMCLMA14b} is readily incorporated
within the same framework. However, it turns
out that the calculation using the Sachs approach is rather redundant,
in a way numerous terms in Eq.~\eqref{eq:III} are cancelled with other
contributions in the final expression. This redundancy may act as an
obstacle in deriving the correct expression of the luminosity distance
at the second order, to which the expression is rather lengthy and involved
(see, e.g., \cite{BAMARI05}).

The Jacobi mapping approach \cite{BODUGA06} has a simple physical
interpretation of its expression of the luminosity distance
in Eqs.~\eqref{eq:JACOBL1} and~\eqref{eq:JACOBL2}, although they have
to be re-arranged in Eq.~\eqref{eq:JACOBL} with the conformally transformed 
metric. As the conformal transformation relates the derivative structures
in Eq.~\eqref{confrel} with two metric tensors in a non-trivial way,
the Jacobi field and its propagation equations also transform in an equally
non-trivial way.
We have identified a properly well-defined Jacobi field $\hat\xi^I$
in the conformally transformed metric, 
with which the form of the propagation equations is preserved as in
the original metric. However, 
the calculation based on the Jacobi mapping results in the
expression similar to the expression obtained in the Sachs approach, 
and hence the redundancy persists in the Jacobi mapping approach.

The geodesic light cone (GLC) approach \cite{GAMAET11} 
is somewhat similar in spirit to the geometric approach, in which the 
observable quantities form the basis. Since it is relatively
new and has been mostly computed in the conformal Newtonian (and recently
in the synchronous gauge \cite{FAGAET13}), further work needs to be done
to ensure the sanity of the formalism, at least, by checking the gauge
invariance of its expressions of the observable quantities.
The great strength of the GLC approach is the simplicity of its expressions
for observable quantities. However, since the GLC approach is adopted with the
full metric, rather than the conformally transformed metric, its calculations
are often more complicated than those in other methods. 

In summary, all these four methods, if properly exercised, 
result in the correct and
consistent expression of the luminosity distance, providing solid theoretical
frameworks. Given the level of consistency with those in the Sachs and 
the geometric approaches, we believe that all the four methods are on equal 
footing and can be readily generalized to compute the higher-order 
corrections in the luminosity distance, 
although the calculations in the Jacobi mapping and the GLC
approaches are performed only with one or two specific gauge conditions.
In particular, the second-order calculations are needed to
compute the mean of the luminosity distance in an inhomogeneous universe,
in which the comparison among other groups is quite difficult.
The unified treatment of the luminosity distance in this work can be
used to go beyond the linear order, providing a crucial way of checking the 
robustness of the calculations and ensuring the consistency of the results 
from different methods.

\acknowledgments
We thank Sang Gyu Biern, Ruth Durrer, Jinn-Ouk Gong, and 
Giuseppe Fanizza for useful discussions. We also thank Gabriele Veneziano
on behalf of the GLC collaboration
for useful comments and clarification in Section~6.
We thank the referee for directing our attention to a different approach
in \cite{SCJE12a}.
We acknowledge support by the Swiss National Science Foundation, and
J.Y. is further supported by
a Consolidator Grant of the European Research Council (ERC-2015-CoG grant
680886).

\bibliography{ms.bbl}

\begin{thebibliography}{38}
\expandafter\ifx\csname natexlab\endcsname\relax\def\natexlab#1{#1}\fi
\expandafter\ifx\csname bibnamefont\endcsname\relax
  \def\bibnamefont#1{#1}\fi
\expandafter\ifx\csname bibfnamefont\endcsname\relax
  \def\bibfnamefont#1{#1}\fi
\expandafter\ifx\csname citenamefont\endcsname\relax
  \def\citenamefont#1{#1}\fi
\expandafter\ifx\csname url\endcsname\relax
  \def\url#1{\texttt{#1}}\fi
\expandafter\ifx\csname urlprefix\endcsname\relax\def\urlprefix{URL }\fi
\providecommand{\bibinfo}[2]{#2}
\providecommand{\eprint}[2][]{\url{#2}}

\bibitem[{\citenamefont{{Sasaki}}(1987)}]{SASAK87}
\bibinfo{author}{\bibfnamefont{M.}~\bibnamefont{{Sasaki}}},
  \bibinfo{journal}{\mnras} \textbf{\bibinfo{volume}{228}},
  \bibinfo{pages}{653} (\bibinfo{year}{1987}).

\bibitem[{\citenamefont{{Umeh} et~al.}(2014{\natexlab{a}})\citenamefont{{Umeh},
  {Clarkson}, and {Maartens}}}]{UMCLMA14a}
\bibinfo{author}{\bibfnamefont{O.}~\bibnamefont{{Umeh}}},
  \bibinfo{author}{\bibfnamefont{C.}~\bibnamefont{{Clarkson}}},
  \bibnamefont{and}
  \bibinfo{author}{\bibfnamefont{R.}~\bibnamefont{{Maartens}}},
  \bibinfo{journal}{Classical and Quantum Gravity}
  \textbf{\bibinfo{volume}{31}}, \bibinfo{pages}{202001}
  (\bibinfo{year}{2014}{\natexlab{a}}), \eprint{1207.2109}.

\bibitem[{\citenamefont{{Bonvin} et~al.}(2006)\citenamefont{{Bonvin}, {Durrer},
  and {Gasparini}}}]{BODUGA06}
\bibinfo{author}{\bibfnamefont{C.}~\bibnamefont{{Bonvin}}},
  \bibinfo{author}{\bibfnamefont{R.}~\bibnamefont{{Durrer}}}, \bibnamefont{and}
  \bibinfo{author}{\bibfnamefont{M.~A.} \bibnamefont{{Gasparini}}},
  \bibinfo{journal}{\prd} \textbf{\bibinfo{volume}{73}},
  \bibinfo{pages}{023523} (\bibinfo{year}{2006}), \eprint{arXiv:0511183}.

\bibitem[{\citenamefont{{Perlmutter} et~al.}(1999)}]{PEADET99}
\bibinfo{author}{\bibfnamefont{S.}~\bibnamefont{{Perlmutter}}}
  \bibnamefont{et~al.}, \bibinfo{journal}{\apj} \textbf{\bibinfo{volume}{517}},
  \bibinfo{pages}{565} (\bibinfo{year}{1999}), \eprint{arXiv:astro-ph/9812133}.

\bibitem[{\citenamefont{{Riess} et~al.}(1998)}]{RIFIET98}
\bibinfo{author}{\bibfnamefont{A.~G.} \bibnamefont{{Riess}}}
  \bibnamefont{et~al.}, \bibinfo{journal}{\aj} \textbf{\bibinfo{volume}{116}},
  \bibinfo{pages}{1009} (\bibinfo{year}{1998}),
  \eprint{arXiv:astro-ph/9805201}.

\bibitem[{\citenamefont{{Jeong} et~al.}(2012)\citenamefont{{Jeong}, {Schmidt},
  and {Hirata}}}]{JESCHI12}
\bibinfo{author}{\bibfnamefont{D.}~\bibnamefont{{Jeong}}},
  \bibinfo{author}{\bibfnamefont{F.}~\bibnamefont{{Schmidt}}},
  \bibnamefont{and} \bibinfo{author}{\bibfnamefont{C.~M.}
  \bibnamefont{{Hirata}}}, \bibinfo{journal}{\prd}
  \textbf{\bibinfo{volume}{85}}, \bibinfo{pages}{023504}
  (\bibinfo{year}{2012}), \eprint{arXiv:1107.5427}.

\bibitem[{\citenamefont{{Yoo}}(2014)}]{YOO14a}
\bibinfo{author}{\bibfnamefont{J.}~\bibnamefont{{Yoo}}},
  \bibinfo{journal}{\cqg} \textbf{\bibinfo{volume}{31}}, \bibinfo{eid}{234001}
  (\bibinfo{year}{2014}), \eprint{arXiv:1409.3223}.

\bibitem[{\citenamefont{{Yoo} and {Zaldarriaga}}(2014)}]{YOZA14}
\bibinfo{author}{\bibfnamefont{J.}~\bibnamefont{{Yoo}}} \bibnamefont{and}
  \bibinfo{author}{\bibfnamefont{M.}~\bibnamefont{{Zaldarriaga}}},
  \bibinfo{journal}{\prd} \textbf{\bibinfo{volume}{90}},
  \bibinfo{pages}{023513} (\bibinfo{year}{2014}), \eprint{1406.4140}.

\bibitem[{\citenamefont{{Gasperini} et~al.}(2011)\citenamefont{{Gasperini},
  {Marozzi}, {Nugier}, and {Veneziano}}}]{GAMAET11}
\bibinfo{author}{\bibfnamefont{M.}~\bibnamefont{{Gasperini}}},
  \bibinfo{author}{\bibfnamefont{G.}~\bibnamefont{{Marozzi}}},
  \bibinfo{author}{\bibfnamefont{F.}~\bibnamefont{{Nugier}}}, \bibnamefont{and}
  \bibinfo{author}{\bibfnamefont{G.}~\bibnamefont{{Veneziano}}},
  \bibinfo{journal}{\jcap} \textbf{\bibinfo{volume}{7}}, \bibinfo{eid}{008}
  (\bibinfo{year}{2011}), \eprint{1104.1167}.

\bibitem[{\citenamefont{{Barausse} et~al.}(2005)\citenamefont{{Barausse},
  {Matarrese}, and {Riotto}}}]{BAMARI05}
\bibinfo{author}{\bibfnamefont{E.}~\bibnamefont{{Barausse}}},
  \bibinfo{author}{\bibfnamefont{S.}~\bibnamefont{{Matarrese}}},
  \bibnamefont{and} \bibinfo{author}{\bibfnamefont{A.}~\bibnamefont{{Riotto}}},
  \bibinfo{journal}{\prd} \textbf{\bibinfo{volume}{71}}, \bibinfo{eid}{063537}
  (\bibinfo{year}{2005}), \eprint{astro-ph/0501152}.

\bibitem[{\citenamefont{{Umeh} et~al.}(2014{\natexlab{b}})\citenamefont{{Umeh},
  {Clarkson}, and {Maartens}}}]{UMCLMA14b}
\bibinfo{author}{\bibfnamefont{O.}~\bibnamefont{{Umeh}}},
  \bibinfo{author}{\bibfnamefont{C.}~\bibnamefont{{Clarkson}}},
  \bibnamefont{and}
  \bibinfo{author}{\bibfnamefont{R.}~\bibnamefont{{Maartens}}},
  \bibinfo{journal}{Classical and Quantum Gravity}
  \textbf{\bibinfo{volume}{31}}, \bibinfo{pages}{205001}
  (\bibinfo{year}{2014}{\natexlab{b}}), \eprint{1402.1933}.

\bibitem[{\citenamefont{{Bertacca} et~al.}(2012)\citenamefont{{Bertacca},
  {Maartens}, {Raccanelli}, and {Clarkson}}}]{BEMAET12}
\bibinfo{author}{\bibfnamefont{D.}~\bibnamefont{{Bertacca}}},
  \bibinfo{author}{\bibfnamefont{R.}~\bibnamefont{{Maartens}}},
  \bibinfo{author}{\bibfnamefont{A.}~\bibnamefont{{Raccanelli}}},
  \bibnamefont{and}
  \bibinfo{author}{\bibfnamefont{C.}~\bibnamefont{{Clarkson}}},
  \bibinfo{journal}{\jcap} \textbf{\bibinfo{volume}{10}}, \bibinfo{pages}{25}
  (\bibinfo{year}{2012}), \eprint{1205.5221}.

\bibitem[{\citenamefont{{Ben-Dayan}
  et~al.}(2013{\natexlab{a}})\citenamefont{{Ben-Dayan}, {Gasperini}, {Marozzi},
  {Nugier}, and {Veneziano}}}]{BEGAET13}
\bibinfo{author}{\bibfnamefont{I.}~\bibnamefont{{Ben-Dayan}}},
  \bibinfo{author}{\bibfnamefont{M.}~\bibnamefont{{Gasperini}}},
  \bibinfo{author}{\bibfnamefont{G.}~\bibnamefont{{Marozzi}}},
  \bibinfo{author}{\bibfnamefont{F.}~\bibnamefont{{Nugier}}}, \bibnamefont{and}
  \bibinfo{author}{\bibfnamefont{G.}~\bibnamefont{{Veneziano}}},
  \bibinfo{journal}{\jcap} \textbf{\bibinfo{volume}{6}}, \bibinfo{eid}{002}
  (\bibinfo{year}{2013}{\natexlab{a}}), \eprint{1302.0740}.

\bibitem[{\citenamefont{{Ben-Dayan}
  et~al.}(2013{\natexlab{b}})\citenamefont{{Ben-Dayan}, {Gasperini}, {Marozzi},
  {Nugier}, and {Veneziano}}}]{BEGAET13a}
\bibinfo{author}{\bibfnamefont{I.}~\bibnamefont{{Ben-Dayan}}},
  \bibinfo{author}{\bibfnamefont{M.}~\bibnamefont{{Gasperini}}},
  \bibinfo{author}{\bibfnamefont{G.}~\bibnamefont{{Marozzi}}},
  \bibinfo{author}{\bibfnamefont{F.}~\bibnamefont{{Nugier}}}, \bibnamefont{and}
  \bibinfo{author}{\bibfnamefont{G.}~\bibnamefont{{Veneziano}}},
  \bibinfo{journal}{\prl} \textbf{\bibinfo{volume}{110}},
  \bibinfo{pages}{021301} (\bibinfo{year}{2013}{\natexlab{b}}),
  \eprint{1207.1286}.

\bibitem[{\citenamefont{{Ben-Dayan} et~al.}(2014)\citenamefont{{Ben-Dayan},
  {Durrer}, {Marozzi}, and {Schwarz}}}]{BEDUET14}
\bibinfo{author}{\bibfnamefont{I.}~\bibnamefont{{Ben-Dayan}}},
  \bibinfo{author}{\bibfnamefont{R.}~\bibnamefont{{Durrer}}},
  \bibinfo{author}{\bibfnamefont{G.}~\bibnamefont{{Marozzi}}},
  \bibnamefont{and} \bibinfo{author}{\bibfnamefont{D.~J.}
  \bibnamefont{{Schwarz}}}, \bibinfo{journal}{\prl}
  \textbf{\bibinfo{volume}{112}}, \bibinfo{pages}{221301}
  (\bibinfo{year}{2014}), \eprint{1401.7973}.

\bibitem[{\citenamefont{{Clarkson} et~al.}(2014)\citenamefont{{Clarkson},
  {Umeh}, {Maartens}, and {Durrer}}}]{CLUMET14}
\bibinfo{author}{\bibfnamefont{C.}~\bibnamefont{{Clarkson}}},
  \bibinfo{author}{\bibfnamefont{O.}~\bibnamefont{{Umeh}}},
  \bibinfo{author}{\bibfnamefont{R.}~\bibnamefont{{Maartens}}},
  \bibnamefont{and} \bibinfo{author}{\bibfnamefont{R.}~\bibnamefont{{Durrer}}},
  \bibinfo{journal}{\jcap} \textbf{\bibinfo{volume}{11}}, \bibinfo{pages}{36}
  (\bibinfo{year}{2014}), \eprint{1405.7860}.

\bibitem[{\citenamefont{{Biern} and {Yoo}}(2016)}]{BIYO16}
\bibinfo{author}{\bibfnamefont{S.~G.} \bibnamefont{{Biern}}} \bibnamefont{and}
  \bibinfo{author}{\bibfnamefont{J.}~\bibnamefont{{Yoo}}},
  \bibinfo{journal}{ArXiv e-prints}  (\bibinfo{year}{2016}),
  \eprint{1606.01910}.

\bibitem[{\citenamefont{{Yoo} et~al.}(2009)\citenamefont{{Yoo}, {Fitzpatrick},
  and {Zaldarriaga}}}]{YOFIZA09}
\bibinfo{author}{\bibfnamefont{J.}~\bibnamefont{{Yoo}}},
  \bibinfo{author}{\bibfnamefont{A.~L.} \bibnamefont{{Fitzpatrick}}},
  \bibnamefont{and}
  \bibinfo{author}{\bibfnamefont{M.}~\bibnamefont{{Zaldarriaga}}},
  \bibinfo{journal}{\prd} \textbf{\bibinfo{volume}{80}},
  \bibinfo{pages}{083514} (\bibinfo{year}{2009}), \eprint{arXiv:0907.0707}.

\bibitem[{\citenamefont{{Yoo}}(2010)}]{YOO10}
\bibinfo{author}{\bibfnamefont{J.}~\bibnamefont{{Yoo}}},
  \bibinfo{journal}{\prd} \textbf{\bibinfo{volume}{82}},
  \bibinfo{pages}{083508} (\bibinfo{year}{2010}), \eprint{arXiv:1009.3021}.

\bibitem[{\citenamefont{{Ade} et~al.}(2014)}]{PLANCK13}
\bibinfo{author}{\bibfnamefont{P.~A.~R.} \bibnamefont{{Ade}}}
  \bibnamefont{et~al.}, \bibinfo{journal}{\aap} \textbf{\bibinfo{volume}{571}},
  \bibinfo{eid}{A16} (\bibinfo{year}{2014}), \eprint{1303.5076}.

\bibitem[{\citenamefont{{Riess} et~al.}(2016)}]{RIMAET16}
\bibinfo{author}{\bibfnamefont{A.~G.} \bibnamefont{{Riess}}}
  \bibnamefont{et~al.}, \bibinfo{journal}{ArXiv e-prints}
  (\bibinfo{year}{2016}), \eprint{1604.01424}.

\bibitem[{\citenamefont{{Fleury} et~al.}(2013)\citenamefont{{Fleury}, {Dupuy},
  and {Uzan}}}]{FLDUUZ13a}
\bibinfo{author}{\bibfnamefont{P.}~\bibnamefont{{Fleury}}},
  \bibinfo{author}{\bibfnamefont{H.}~\bibnamefont{{Dupuy}}}, \bibnamefont{and}
  \bibinfo{author}{\bibfnamefont{J.-P.} \bibnamefont{{Uzan}}},
  \bibinfo{journal}{Physical Review Letters} \textbf{\bibinfo{volume}{111}},
  \bibinfo{pages}{091302} (\bibinfo{year}{2013}), \eprint{1304.7791}.

\bibitem[{\citenamefont{{Sachs} and {Wolfe}}(1967)}]{SAWO67}
\bibinfo{author}{\bibfnamefont{R.~K.} \bibnamefont{{Sachs}}} \bibnamefont{and}
  \bibinfo{author}{\bibfnamefont{A.~M.} \bibnamefont{{Wolfe}}},
  \bibinfo{journal}{\apj} \textbf{\bibinfo{volume}{147}}, \bibinfo{pages}{73}
  (\bibinfo{year}{1967}).

\bibitem[{\citenamefont{{Blandford} and {Narayan}}(1986)}]{BLNA86}
\bibinfo{author}{\bibfnamefont{R.}~\bibnamefont{{Blandford}}} \bibnamefont{and}
  \bibinfo{author}{\bibfnamefont{R.}~\bibnamefont{{Narayan}}},
  \bibinfo{journal}{\apj} \textbf{\bibinfo{volume}{310}}, \bibinfo{pages}{568}
  (\bibinfo{year}{1986}).

\bibitem[{\citenamefont{{Bartelmann} and {Schneider}}(2001)}]{BASC01}
\bibinfo{author}{\bibfnamefont{M.}~\bibnamefont{{Bartelmann}}}
  \bibnamefont{and}
  \bibinfo{author}{\bibfnamefont{P.}~\bibnamefont{{Schneider}}},
  \bibinfo{journal}{\physrep} \textbf{\bibinfo{volume}{340}},
  \bibinfo{pages}{291} (\bibinfo{year}{2001}), \eprint{arXiv:9912508}.

\bibitem[{\citenamefont{{Ben-Dayan}
  et~al.}(2012{\natexlab{a}})\citenamefont{{Ben-Dayan}, {Gasperini}, {Marozzi},
  {Nugier}, and {Veneziano}}}]{BEGAET12a}
\bibinfo{author}{\bibfnamefont{I.}~\bibnamefont{{Ben-Dayan}}},
  \bibinfo{author}{\bibfnamefont{M.}~\bibnamefont{{Gasperini}}},
  \bibinfo{author}{\bibfnamefont{G.}~\bibnamefont{{Marozzi}}},
  \bibinfo{author}{\bibfnamefont{F.}~\bibnamefont{{Nugier}}}, \bibnamefont{and}
  \bibinfo{author}{\bibfnamefont{G.}~\bibnamefont{{Veneziano}}},
  \bibinfo{journal}{\jcap} \textbf{\bibinfo{volume}{4}}, \bibinfo{eid}{036}
  (\bibinfo{year}{2012}{\natexlab{a}}), \eprint{1202.1247}.

\bibitem[{\citenamefont{{Ben-Dayan}
  et~al.}(2012{\natexlab{b}})\citenamefont{{Ben-Dayan}, {Marozzi}, {Nugier},
  and {Veneziano}}}]{BEMAET12a}
\bibinfo{author}{\bibfnamefont{I.}~\bibnamefont{{Ben-Dayan}}},
  \bibinfo{author}{\bibfnamefont{G.}~\bibnamefont{{Marozzi}}},
  \bibinfo{author}{\bibfnamefont{F.}~\bibnamefont{{Nugier}}}, \bibnamefont{and}
  \bibinfo{author}{\bibfnamefont{G.}~\bibnamefont{{Veneziano}}},
  \bibinfo{journal}{\jcap} \textbf{\bibinfo{volume}{11}}, \bibinfo{eid}{045}
  (\bibinfo{year}{2012}{\natexlab{b}}), \eprint{1209.4326}.

\bibitem[{\citenamefont{{Kaiser} and {Hudson}}(2015)}]{KAHU15a}
\bibinfo{author}{\bibfnamefont{N.}~\bibnamefont{{Kaiser}}} \bibnamefont{and}
  \bibinfo{author}{\bibfnamefont{M.~J.} \bibnamefont{{Hudson}}},
  \bibinfo{journal}{\mnras} \textbf{\bibinfo{volume}{450}},
  \bibinfo{pages}{883} (\bibinfo{year}{2015}), \eprint{1411.6339}.

\bibitem[{\citenamefont{{Schmidt} and {Jeong}}(2012)}]{SCJE12a}
\bibinfo{author}{\bibfnamefont{F.}~\bibnamefont{{Schmidt}}} \bibnamefont{and}
  \bibinfo{author}{\bibfnamefont{D.}~\bibnamefont{{Jeong}}},
  \bibinfo{journal}{\prd} \textbf{\bibinfo{volume}{86}}, \bibinfo{eid}{083527}
  (\bibinfo{year}{2012}), \eprint{1204.3625}.

\bibitem[{\citenamefont{{Jeong} and {Schmidt}}(2014)}]{JESC13}
\bibinfo{author}{\bibfnamefont{D.}~\bibnamefont{{Jeong}}} \bibnamefont{and}
  \bibinfo{author}{\bibfnamefont{F.}~\bibnamefont{{Schmidt}}},
  \bibinfo{journal}{\prd} \textbf{\bibinfo{volume}{89}},
  \bibinfo{pages}{043519} (\bibinfo{year}{2014}), \eprint{1305.1299}.

\bibitem[{\citenamefont{{Ehlers}}(1961)}]{EHLER61}
\bibinfo{author}{\bibfnamefont{J.}~\bibnamefont{{Ehlers}}},
  \emph{\bibinfo{title}{{Proceedings of the mathematical-natural science of the
  Mainz academy of science and literature, translated in Gen. Rel. Grav. 25,
  1225, 1993}}}, vol. \bibinfo{volume}{792} (\bibinfo{year}{1961}).

\bibitem[{\citenamefont{{Ellis}}(1971)}]{ELLIS71}
\bibinfo{author}{\bibfnamefont{G.~F.~R.} \bibnamefont{{Ellis}}}, in
  \emph{\bibinfo{booktitle}{General Relativity and Cosmology}}, edited by
  \bibinfo{editor}{\bibnamefont{{R.~K.~Sachs}}} (\bibinfo{year}{1971}), pp.
  \bibinfo{pages}{104--182}.

\bibitem[{\citenamefont{{Sachs}}(1961)}]{SACHS61}
\bibinfo{author}{\bibfnamefont{R.}~\bibnamefont{{Sachs}}},
  \bibinfo{journal}{Royal Society of London Proceedings Series A}
  \textbf{\bibinfo{volume}{264}}, \bibinfo{pages}{309} (\bibinfo{year}{1961}).

\bibitem[{\citenamefont{{Kristian} and {Sachs}}(1966)}]{KRSA66}
\bibinfo{author}{\bibfnamefont{J.}~\bibnamefont{{Kristian}}} \bibnamefont{and}
  \bibinfo{author}{\bibfnamefont{R.~K.} \bibnamefont{{Sachs}}},
  \bibinfo{journal}{\apj} \textbf{\bibinfo{volume}{143}}, \bibinfo{pages}{379}
  (\bibinfo{year}{1966}).

\bibitem[{\citenamefont{{Seitz} et~al.}(1994)\citenamefont{{Seitz},
  {Schneider}, and {Ehlers}}}]{SESCEH94}
\bibinfo{author}{\bibfnamefont{S.}~\bibnamefont{{Seitz}}},
  \bibinfo{author}{\bibfnamefont{P.}~\bibnamefont{{Schneider}}},
  \bibnamefont{and} \bibinfo{author}{\bibfnamefont{J.}~\bibnamefont{{Ehlers}}},
  \bibinfo{journal}{Classical and Quantum Gravity}
  \textbf{\bibinfo{volume}{11}}, \bibinfo{pages}{2345} (\bibinfo{year}{1994}),
  \eprint{arXiv:9403056}.

\bibitem[{\citenamefont{{Lewis} and {Challinor}}(2006)}]{LECH06}
\bibinfo{author}{\bibfnamefont{A.}~\bibnamefont{{Lewis}}} \bibnamefont{and}
  \bibinfo{author}{\bibfnamefont{A.}~\bibnamefont{{Challinor}}},
  \bibinfo{journal}{\physrep} \textbf{\bibinfo{volume}{429}},
  \bibinfo{pages}{1} (\bibinfo{year}{2006}), \eprint{arXiv:astro-ph/0601594}.

\bibitem[{\citenamefont{{Bernardeau} et~al.}(2012)\citenamefont{{Bernardeau},
  {Bonvin}, {Van de Rijt}, and {Vernizzi}}}]{BEBOET12}
\bibinfo{author}{\bibfnamefont{F.}~\bibnamefont{{Bernardeau}}},
  \bibinfo{author}{\bibfnamefont{C.}~\bibnamefont{{Bonvin}}},
  \bibinfo{author}{\bibfnamefont{N.}~\bibnamefont{{Van de Rijt}}},
  \bibnamefont{and}
  \bibinfo{author}{\bibfnamefont{F.}~\bibnamefont{{Vernizzi}}},
  \bibinfo{journal}{\prd} \textbf{\bibinfo{volume}{86}},
  \bibinfo{pages}{023001} (\bibinfo{year}{2012}), \eprint{1112.4430}.

\bibitem[{\citenamefont{{Fanizza} et~al.}(2013)\citenamefont{{Fanizza},
  {Gasperini}, {Marozzi}, and {Veneziano}}}]{FAGAET13}
\bibinfo{author}{\bibfnamefont{G.}~\bibnamefont{{Fanizza}}},
  \bibinfo{author}{\bibfnamefont{M.}~\bibnamefont{{Gasperini}}},
  \bibinfo{author}{\bibfnamefont{G.}~\bibnamefont{{Marozzi}}},
  \bibnamefont{and}
  \bibinfo{author}{\bibfnamefont{G.}~\bibnamefont{{Veneziano}}},
  \bibinfo{journal}{\jcap} \textbf{\bibinfo{volume}{11}}, \bibinfo{eid}{019}
  (\bibinfo{year}{2013}), \eprint{1308.4935}.

\end{thebibliography}

\end{document}